\documentclass{article}
\usepackage{fullpage}
\usepackage{amsfonts,amssymb}
\usepackage{amsthm}
\usepackage{tikz}
\usetikzlibrary{calc}
\usetikzlibrary{arrows}
\usepackage{tikz-3dplot}
\usepackage{color} 
\usepackage[fleqn]{amsmath}
\usepackage{amsthm}
\usepackage{amssymb}
\usepackage{amsfonts}
\usepackage{bbm}
\usepackage{epsfig}
\usepackage{times}
\usepackage{hyperref}
\usepackage{bm}
\usepackage{float} 
\usepackage{appendix} 
\usepackage{lscape}
\usepackage{cite}
\usepackage{mathrsfs}
\usepackage{appendix}
\usepackage{setspace}
\usepackage{mathtools}
\usepackage{authblk} 

\theoremstyle{definition}
\newtheorem{definition}{Definition}[section]
\begin{document}

\title{Tabletop Experiments for Quantum Gravity Are Also Tests of the Interpretation of Quantum Mechanics}

 \author{Emily Adlam  \thanks{The Rotman Institute of Philosophy, 1151 Richmond Street, London N6A5B7 \texttt{eadlam90@gmail.com} }}

\maketitle

Recently there has been a great deal of interest in tabletop experiments  intended to exhibit the quantum nature of gravity by demonstrating that it can induce entanglement. In order to evaluate the significance of these experiments, we must determine if there is  any interesting class of possibilities that will be convincingly ruled out if it turns out that gravity can indeed induce entanglement. In this article, we suggest that this result would rule out a class of quantum gravity models that we refer to as  $\psi$-incomplete quantum gravity (PIQG) - i.e. models  of the interaction between quantum mechanics and gravity in which gravity is coupled to non-quantum beables rather than quantum beables. This indicates that the results of the tabletop experiments can also be understood as providing new information about the correct interpretation of quantum mechanics. 

In particular, a major motivation for these experiments is the idea that they   witness the existence of superpositions of spacetimes, and thus it should be emphasised that claims about the existence of spacetime superpositions are not interpretation-neutral.  $\psi$-complete interpretations of quantum mechanics, like the Everett interpretation, almost universally tell us that spacetime superpositions are possible, whereas  in $\psi$-incomplete, $\psi$-nonphysical interpretations it seems more natural to predict that spacetime superpositions are not possible. Meanwhile $\psi$-incomplete, $\psi$-supplemented interpretations present us with a more complex picture where we may or may not end up predicting that spacetime superpositions are possible, depending on the  way in which the coupling between spacetime and matter  is constructed. Thus roughly speaking, a positive result to the tabletop experiments should increase our confidence in $\psi$-complete interpretations, whilst a negative result should instead increase our confidence in $\psi$-incomplete interpretations.
  
In section  \ref{intro} we introduce PIQG models, and then in section \ref{inference} we make the reasoning more precise  by presenting   a set of inferences that may be made about the ontology of quantum mechanics based on the results of tabletop experiments. In section \ref{PIQG} we discuss some existing PIQG models and consider what more needs to be done to make these sorts of approaches more appealing. There are two competing paradigms for the interpretation of these experiments, which have been dubbed the `Newtonian' paradigm and the `tripartite' paradigm: here we largely work within the tripartite paradigm, because the tripartite view is specifically concerned with ontological aspects of the tabletop experiments and that makes it a suitable setting for enquiries about the ontology of quantum mechanics, but in section \ref{changegr} we consider what conclusions can be drawn if one does not presuppose the tripartite view. Finally in section \ref{cosmology}  we discuss a cosmological phenomenon which could be regarded as providing evidence for PIQG models.

\section{Introduction \label{intro}}

Ref \cite{Craig2001-CRAPMP} notes that debates over the need to quantize gravity can be separated into two parts: `\emph{The first problem can be expressed by asking “why do we need a quantum theory of gravity at all?” and the second, assuming that the first one was answered in the affirmative, by inquiring “why do we have to quantize gravity for the purpose of finding a quantum theory of gravity?}
In this article we will take for granted that we do need a quantum theory of gravity (i.e. a theory which explains how gravity and quantum mechanics are related) and address ourselves wholly to the second question. However, our interest here is in matters of interpretation rather than methodology, so we will emphasize a more specific ontological question: `Can there be superpositions of spacetimes?' 
Obviously these points are closely linked: all extant attempts to quantize gravity predict superpositions of spacetimes, and conversely, if there can indeed be superpositions of spacetimes it seems natural to expect that the gravitational field must be quantized.  

At present  we have no direct empirical evidence showing that spacetime can be put in superpositions. What we do have is evidence for a number of applications of low-energy quantum gravity, which is `\emph{the straightforward conservative theory one obtains by treating GR as an effective field theory, quantized through path-integral means}'\cite{wallace2021quantum}. Low-energy quantum gravity leads to a semiclassical approximation which has been applied successfully throughout astrophysics and cosmology, and its application in the perturbative incoherent regime is responsible for the models of inflation which correctly predict the  spectrum of CMB fluctuations. It also predicts, correctly, that  quantum systems feel the effect of gravity (for example, a  neutron interferometer has been used to demonstrate that the gravitational field affects the behaviour of
quantum systems such as a beam of neutrons\cite{10.2307/24966324}) and that  quantum systems can be sources of gravitational fields (as in the Page-Geilker experiments\cite{PhysRevLett.47.979}). Indeed, if one accepts that all matter is made up of quantum systems, then really any demonstration of gravitating matter can be regarded as a demonstration that quantum systems both feel the effects of and source gravitational fields. But Wallace observes that `\emph{At present, we have little evidence supporting applications of LEQG in the perturbative/coherent regime ... it is this regime which distinguishes unitary quantum gravity from variants in which superpositions of mass distributions cause wavefunction collapse.}'\cite{wallace2021quantum}. That is to say, we have not yet made any observations of the regime in which superpositions of spacetimes become important.

Nonetheless, the  existence of spacetime superpositions is expected by many physicists, and is predicted by most mainstream approaches to quantum gravity\cite{Blau2009,cc}. This is   particularly clear  in  the context of recent proposals for tabletop experiments probing quantum gravity.  For example, in the proposed Bose-Marletto-Vedral (BMV) experiment\cite{2017sew,2017gieb}, experimenters will prepare two particles of mass $m$ at a small distance $d$ from one another for a time $t$. The gravitational effect of one particle on the other will produce a phase shift $\delta \phi = \frac{G m^2 t}{\hbar d}$. Each particle will be prepared in a superposition of two different spatial positions, giving rise to four different configurations corresponding to four different branches of the wavefunction, with different phase changes in different branches. The particles are thus expected to become entangled, and experimenters will check for the presence of entanglement  by looking for Bell correlations in subsequent spin measurements. As argued in ref \cite{2017gieb}, if gravity can entangle two systems then it must have some non-commuting variables, so if entanglement is detected in this experiment we will be able to conclude that gravity is indeed quantum. Moreover, the standard analysis assumes there will be different spacetime structure in each of the four branches of the wavefunction, and it is this superposition of spacetimes which will mediate the different phase changes in each branch. 
   
Let us  try to understand why so many physicists are so confident that there can indeed be spacetime superpositions in scenarios like that of the BMV experiment. There exist a variety of formal arguments aiming to show that gravity must be quantized, which we will address in section \ref{mustbe}, but as noted by Callender and Huggett, `\emph{most physicists arguing for the necessity of quantum gravity do not take (these arguments) as the main reason for quantizing the gravitational field. Rather, they usually point to a list of what one might call methodological points in favour of quantum gravity}.' Similarly Mattingly argues that the attempt to quantize gravity often seems driven by a commitment to unification rather than any specific theoretical or empirical arguments\cite{inbookmatt}  - as for example in Rovelli's comment `\emph{We have learned from QM that every dynamical object has quantum properties, which can be captured by appropriately formulating its dynamical theory within the general scheme of QM.}'\cite{Rovelli2001-ROVQSW} As can be seen from this quote, the relevant notion of unification here is the idea that  there should be ultimately only one type of stuff  - i.e. quantum stuff - and therefore since ordinary matter gravitates, and ordinary matter is assumed to be composed entirely of quantum stuff, it follows that quantum stuff must gravitate. But not all interpretations of quantum mechanics agree that there exists only `quantum stuff,' and therefore this conclusion is by no means inevitable. 
 
In this article, we will use the term `the quantum sector' to refer to any beables    which possess noncommuting variables, following ref \cite{marletto2017need} in taking this to be the defining feature of quantum behaviour: `\emph{This is precisely what one means by the field being “quantum”: it must have at least two non-commuting observables.}'  In standard non-relativistic quantum mechanics the quantum sector is just the quantum state or wavefunction, but we employ this more general term in order to encompass other possible quantum descriptions - for example quantum fields,  wavefunctionals and so on. The nonquantum sector meanwhile includes all beables which do not possess noncommuting variables, such as de Broglie Bohm particle positions\cite{Passon2006-PASWYA,holland1995quantum,SEPBohm},  `flashes' in the Bell flash ontology\cite{Bell1985,Gisin2013}, single-valued mass or mass-energy distributions\cite{Allori_2013} and so on. Note that a beable has noncommuting variables if and only if it can be in superposition states,  and therefore the quantum sector  can participate in superpositions while the nonquantum sector cannot.  In the standard language of quantum foundations\cite{adlam_2021}, an interpretation of quantum mechanics whose ontology includes only the quantum sector is known as a $\psi$-complete interpretation, while an interpretation which includes a physically real nonquantum sector is known as a $\psi$-incomplete interpretation. If the  interpretation tells us that  both the quantum sector and the nonquantum sector are physically real, we call it $\psi$-supplemented, whereas if it tells us that only the nonquantum sector is physically real  it is usually described as $\psi$-epistemic. However, we will not use the term $\psi$-epistemic in this article, as we do not wish to assume that  whenever the quantum state is not an element of physical reality it is necessarily just a description of knowledge - for example, there have been proposals that it should be regarded as a modal fact\cite{sep-qm-modal} or a  feature of the laws of nature\cite{inbooksahc}\cite[p. 10--11]{Durr}. So we will instead refer to such models as $\psi$-nonphysical.  We observe that although these terms are typically applied in the framework of non-relativistic quantum mechanics,   one can easily extend them to a field-theoretic context: a $\psi$-incomplete  approach to quantum field theory has an ontology which includes something more than just the quantum fields, and a $\psi$-nonphysical approach has an ontology which does not include quantum fields as an element of reality (though again, they might be regarded as encoding modal facts or laws of nature). \footnote{There exist some interpretations of quantum mechanics which don't fit easily into this taxonomy. In particular, it's difficult to define a clear ontology for antirealist or instrumentalist approaches, as well as interpretations which relativize their descriptions to an observer, such as the   neo-Copenhagen interpretations, QBism, and relational quantum mechanics. In this article we will confine our attention to interpretations which can be unambiguously classified as $\psi$-complete or $\psi$-incomplete, though it would be interesting to address some of these more complex cases in future work.}

For our purposes, the most important feature of $\psi$-incomplete models is that they typically have a story to tell about matter which differs from the common view that matter is simply composed of quantum fields. Thus in such a model we don't necessarily have to say that the quantum sector carries energy and mass, since the energy and mass could instead be carried by the nonquantum sector.  Indeed, in a $\psi$-nonphysical model quantum fields certainly cannot carry energy and mass, since they are not elements of physical reality. Of course, in most $\psi$-incomplete models the quantum sector plays an important role in dictating the behaviour of the nonquantum sector, and thus in many cases  it may \emph{look} a lot like the fields themselves carry energy and mass, but that could  simply be a misinterpretation. This means that in the context of a $\psi$-incomplete interpretation we can consistently maintain that the stress-energy tensor  does not in fact supervene on the quantum sector, but only on the nonquantum sector, meaning that the stress-energy tensor will not enter into superpositions. And therefore since it is the stress-energy tensor which determines the structure of spacetime via the Einstein Field Equations, the structure of spacetime will not enter into superpositions either. Thus the conclusion that spacetime superpositions can  exist implicitly depends on pre-existing beliefs about the interpretation of quantum mechanics, so for those who remain open-minded about the interpretation of quantum mechanics, the existence of spacetime superpositions should not be a foregone conclusion.

For example, take the Bell flash ontology, which tells us that matter is not a continuous, persisting substance; rather it is a  `constellation of flashes' and nothing whatsoever exists in between the flashes\cite{Bell1985,Gisin2013}. The Bell flash ontology is $\psi$-incomplete, since the flashes have no noncommuting variables, and $\psi$-nonphysical, since the quantum state is not regarded as being part of physical reality in this picture - it is simply a feature of the laws of nature which govern the distribution of flashes.  It is therefore natural to suppose that in the Bell flash ontology the  stress-energy tensor should supervene on the distribution of the flashes and not on quantum states or fields. Thus, since the flashes cannot enter into superpositions, nor can the stress-energy tensor or the structure of spacetime.  In particular, the flash ontology tells us that a spacetime region which according to the standard description contains a mass in `superposition' is in fact completely empty: there will be some flashes in the spacetime region of the state preparation, and also flashes in the spacetime region of the measurement at the end of the experiment, but if we succeed in keeping the mass in superposition in the intervening period, there will be no flashes in between the preparation and measurement. Thus there is no mass or energy in that region to  gravitate, and therefore from the point of view of the  flash ontology there is no reason to expect that a superposition of masses will give rise to a spacetime superposition.  
  
Conversely, in the context of interpretations which are $\psi$-complete, such as the Everett interpretation\cite{Wallace}, all matter is made of quantum fields, and therefore it would seem quite difficult to come up with versions of these interpretations which do not predict the existence of spacetime superpositions. For $\psi$-complete approaches entail that a mass in a spatial superposition is exactly the same kind of thing as a mass which is not in a spatial superposition, and there would seem to be no natural way  to insist that only parts of the quantum state which are sufficiently well-localised can be sources of gravitational fields -   since no quantum state is ever perfectly localised, there would be no clear line between gravitating and non-gravitating matter within such a picture.  Also, according to the Everett interpretation  we routinely end up with superpositions of macroscopically different states of affairs, including  different arrangements of very large objects like planets, and if we believe that general relativity is at least roughly correct in its usual domain of application, such different states of affairs must certainly be associated with different configurations of spacetime. Thus the Everettians seem to have little choice but to accept the existence of spacetime superpositions. So from the Everettian   point of view  it stands to reason that tabletop experiments aiming to demonstrate the existence of spacetime superpositions will eventually succeed, which means that the failure of such experiments would be a blow to the Everett interpretation. Of course there are probably ways that the Everett interpretation could   be adapted to deal with this turn of events, but it is certainly not what Everettians would most naturally expect.

 \section{What can we infer from the results of tabletop experiments? \label{inference}}

 We should begin by acknowledging that there remains some controversy about whether the BMV experiment really demonstrates that gravity is quantum, not least because there is some disagreement around what it actually means for gravity to be quantum. As discussed in ref \cite[p. 21--25]{hugforth} there are two opposing schools of thought. On the one hand, in the `tripartite' paradigm it is assumed that the gravitational interaction between the superposed particles must be mediated, and then it is shown that any mediator which can enduce entanglement must indeed be quantum, in the sense that it must possess commuting variables\cite{2017gieb,PhysRevD.102.086012}. On the other hand, in the `Newtonian' paradigm  it is shown that applying constraint quantisation to an approximation known as linearized gravity yields the conclusion that only the Newtonian part of the GR action is relevant to the description of the BMV experiment, and thus the interaction between the superposed particles involves only a gauge constraint rather than a dynamically propagating degree of freedom\cite{anastopoulos2018comment}; then it is argued that  `\emph{according to GR, the two parts of a quantum bipartite system that interact gravitationally in the Newtonian regime do so without an intermediate degree of freedom}'\cite{Anastopoulos_2021} and therefore the experiment does not show that any gravitational degrees of freedom must be quantum.
 
 These two approaches   have different starting assumptions and take very different stances on the appropriate way to draw conclusions in a scenario where we are still unsure about the underlying fundamental theory.  The Newtonian approach begins by assuming that the BMV experiment is appropriately described by the mathematical tools of quantum field theory together with a linearized gravity approximation, and then tries to draw conclusions from the details of the resulting perturbative expansion, without making any commitments regarding the nature of the underlying  ontology (the perturbative expansion is after all an approximation technique rather than a mathematical model which one might take to directly represent some element of the ontology). The success of field theoretic methods in low energy quantum gravity provides some justification for the choice to  represent gravity as a quantum field theory in this analysis, but there are reasons to proceed with caution, as one might perhaps feel that   the analysis of an experiment explicitly intended to establish whether or not gravity is quantum  should not lean too strongly on the assumption that gravity can be modelled using the tools of quantum field theory. In particular, ref  \cite[p. 23]{hugforth} notes that since there is still considerable uncertainty around what the final theory of gravity will look like, perhaps we should not attach too much weight to the   technical distinction between gauge degrees of freedom and `true' degrees of freedom. After all, a similar argument shows that time itself becomes  `pure gauge' under constraint quantisation, and the most popular response to this so-called `problem of time' is to observe that time can be gauge at the level of the constraint description whilst still representing an important feature of reality  from the point of view of  observers inside the universe\cite{articleIsh,Bojowald_2011,2021trinity};  and in a similar way,   even if the  gravitational interaction in the BMV regime is solely derived from a gauge degree of freedom, that gauge degree of freedom could still represent a real feature of the structure of spacetime from the point of view of observers inside the universe such as ourselves, and therefore  the claim that only gauge degrees of freedom are involved in the BMV experiment is not necessarily in contradiction with the idea that a positive result to the BMV experiment would demonstrate a superposition of spacetimes.

The tripartite paradigm, on the other hand, avoids leaning on specific mathematical details and instead  attempts to come to grips with the underlying ontology of the BMV experiment. Rovelli, working  within the tripartite paradigm, explains the rationale behind the approach as follows: `\emph{the knowledge that gravity is mediated by a field (in fact, a relativistic field) is needed for the interpretation of the experiment. If gravity was an instantaneous action at a distance and not mediated by a field, then we could not conclude anything from the experiment itself}''\cite{2019otpo}.  That is to say, the tripartite paradigm interprets GR as making an ontological claim about the nature of the gravitational interaction and assumes that this claim is  correct at the scales relevant for the BMV experiment. The analysis of the experiment is then relatively straightforward: a successful result for the BMV experiment would demonstrate that masses  in superposition continue to source gravitational fields while they are superposed, and hence a   superposition of mass distributions must give rise to a superposition of gravitational fields, which shows that gravity must be quantized, in the sense that the gravitational field must have non-commuting variables. Questions about the correctness of the tripartite analysis thus come down to deciding   whether  or not it is permissible to assume that GR's account of gravity as mediated by a field (or something similar to a field) is  valid in the setting of the BMV experiments\footnote{Of course one might also object to the tripartite analysis on antirealist or instrumentalist grounds, i.e. by criticizing its aim of coming to grips with the ontology underlying the BMV experiment; but this type of objection will not concern us here, because we are interested in drawing conclusions about the ontology of quantum mechanics, and hence we are already committed to a realist, ontological account.}. One might perhaps oppose the  tripartite paradigm on the grounds that  several current approaches to quantum gravity tell us that   spacetime and gravity are only  emergent, high-level features of reality\cite{LinnemannVisser}, so we cannot be sure that the high-level general relativistic account of them is still correct at the scales relevant to the BMV experiment. Viewed in this way, the key assumption made by the tripartite analysis is that, if  the general relativistic account of spacetime breaks down in a significant way at some scale, it does so at a scale lower than the BMV experiment scales.  Different approaches to quantum gravity, and different interpretations of quantum mechanics, will say different things about whether the general-relativistic account breaks down and the scales at which it breaks down, and thus there is a sense in which attempts to use the BMV experiment to learn something about quantum gravity and/or the interpretation of quantum mechanics are somewhat circular. 

However, we  believe that the tripartite paradigm is worth taking seriously for a number of reasons. First, various theoretical arguments provide  justification for thinking that GR's account of spacetime should indeed still hold at the BMV scales - for example, it's often thought that the scale at which spacetime emerges in quantum gravity is around the Planck scale, in which case  the scales relevant to the BMV experiment are comfortably in the regime in which we'd expect the general relativistic account of spacetime to be appropriate. Moreover most existing approaches to quantum gravity, including string theory and loop quantum gravity, appear to agree with this conclusion.  Second, because part of the definition of PIQG models specifies that GR's account of spacetime is correct at the BMV scales, evidently the  tripartite analysis is appropriate insofar as our aim is to argue that a convincing positive result  to the tabletop experiments would rule out  PIQG models: for if it is not in fact true that GR's  account of spacetime is valid at the BMV scales, then PIQG models are wrong in any case, so from that point of view we lose nothing by assuming that GR's account of spacetime is correct at the BMV scales. Finally,  the  tripartite analysis   allows us to treat the tabletop experiments as a window into the ontology of quantum mechanics, thus providing the rare opportunity to obtain new empirical evidence which bears directly on the interpretation of quantum mechanics, and we consider that it would be a mistake to refrain from taking this opportunity simply because the analysis rests on assumptions which are not fully secure. Thus we  consider that the appropriate course of action is to analyse the BMV experiment   in the context of the   tripartite paradigm but also keep an open mind to possibilities outside that paradigm: thus  in this section, we consider what can be be inferred about the interpretation of quantum mechanics from  the results of tabletop experiments  within the  tripartite paradigm, while  in section \ref{changegr} we consider some approaches beyond the   tripartite paradigm. Because the claim that  the BMV experiment is a probe of quantum gravity depends on the tripartite paradigm,   the results of this section  demonstrate that,  insofar as the BMV experiment can be regarded as a probe of quantum gravity, it is also a test of the interpretation of   quantum mechanics.
 
Given that we are particularly interested in using the BMV experiment to determine whether or not there can be superpositions of \emph{spacetimes} (rather than just gravitational fields or other gravitational mediators), we will sometimes have reason to adopt a stronger version of the tripartite paradigm. As presented in refs \cite{2017gieb,PhysRevD.102.086012} the tripartite paradigm assumes only  that the gravitational interaction is locally mediated, without saying anything about the relationship between spacetime and gravity. Of course it is possible to imagine models in which gravity is locally mediated but the gravitational field is not identified with  spacetime structure at the scales of the BMV experiment, and we discuss this possibility in section  \ref{changegr}, but any model of this kind comes at the cost of undermining GR's highly appealing unification of gravity and spacetime, and thus if one is going to lean on ideas from GR to interpret the BMV experiment, one may prefer to import GR's entire account of gravity rather than selecting parts of it in an ad hoc way. In addition, if one makes only very minimal assumptions about the nature of gravity in one's analysis of the BMV experiment,  one is able to conclude that `gravity is quantised' only in a very weak sense which may not translate readily to any conclusions about the thicker concept of gravity that appears in classical general relativity.  Therefore there are some good reasons to adopt a stronger tripartite paradigm  which includes commitments to all of  the  key  features of GR's account of spacetime - in particular, a commitment  to the claim that at the scales relevant for the BMV experiments the gravitational field  is to be identified with the metric structure of spacetime, which means that if the BMV experiment demonstrates a superposition of gravitational fields it also demonstrates a superposition of spacetimes.  This is in fact the route taken  by Rovelli, who writes, `\emph{Since the gravitational field is the geometry of spacetime (measured by rods and clocks), the BMV effect counts as evidence that quantum superposition of different spacetime geometries is possible, can be achieved in the lab, and has observable effects}'\cite{2019otpo}. That said, we will not need the strong tripartite paradigm for our first few inferences, so to begin with we will work   within the weak tripartite paradigm, moving to the strong paradigm only when we wish to make some stronger inferences.

Thus let us consider an experiment like the BMV proposal in which some sufficiently massive objects (e.g. nanoparticles) are placed in a superposition of different spatial locations. We will say that this experiment has a positive result if we are able to use the setup to violate a Bell's inequality and thus demonstrate that gravity can generate entanglement; we will say the result is \emph{convincing} if it is stable and repeatable such that we are eventually convinced we are  witnessing a real effect and not merely an experimental anomaly. Thus we argue that, subject to the assumption that gravity is mediated at the BMV scales, the following four inferences can be made about this scenario.  It should be reinforced that  the arrows here denote `it is reasonable to think,' rather than deductive inference - given sufficient effort one could probably come up with exotic models which would get around the arguments made here, but nonetheless we consider that  all of these inferences have a reasonable degree of plausibility.   
  
 \begin{enumerate} 
 \item The quantum  description of this scenario is complete $\rightarrow$   the BMV experiment will have a  convincing positive result
\item The BMV experiment has a convincing positive result $\rightarrow$ the quantum sector is physically real
\item  The BMV experiment has a convincing negative result  $\rightarrow$ the quantum description of this scenario is not complete 
 \item The quantum sector is not physically real $\rightarrow$ the BMV experiment will have a convincing negative result
 
\end{enumerate} 

The following two inferences are also possible, although they require the strong tripartite paradigm and some further assumptions:

 \begin{enumerate}  
  \setcounter{enumi}{4} 
\item  The BMV experiment has a convincing positive result   $\rightarrow$ the quantum description of this scenario is  complete
\item The BMV experiment has a convincing negative result   $\rightarrow$ the quantum sector is not physically real
 
\end{enumerate}

To explain these inferences, we will use the following conventions.  Let $G$ be the gravitational field (or some other entity which is responsible for mediating the gravitational interaction; for simplicity we will continue to use the term `gravitational field' but for the first four inferences our arguments do not depend on any details of the nature of $G$). Let  $Q$ be the quantum sector, and $C$ the nonquantum sector. By definition $Q$ has at least two variables $\{ q_1, q_2\}$ which do not commute, which is to say that when $Q$ is found in an eigenstate of $q_1$, it is in a superposition of eigenstates of $q_2$, and vice versa. We will take it that eigenstates of $q_1$ correspond to classical mass-energy distributions, so eigenstates of $q_2$ represent superpositions of classical mass-energy distributions. By definition $C$ has no noncommuting variables so we can combine all its variables into a single variable $c_1$. We will take it that all the possible states of $C$ correspond to classical mass-energy distributions, so the  variable $c_1$ commutes with $q_1$. 

We will suppose that $G$ has at least one variable $g_1$, and possibly also a second variable $g_2$ which does not commute with $g_1$. We will take it that  eigenstates of $g_1$ correspond to  classical configurations of the gravitational field (or other mediating entity), so if there exists such a variable $g_2$ then its eigenstates correspond to superpositions of classical configurations of the gravitational field; that is, $G$ has a variable $g_2$ which does not commute with $g_1$ if and only if superpositions of gravitational fields are possible. Invoking the arguments of refs \cite{2017gieb,PhysRevD.102.086012}, we infer that since we are presupposing the tripartite paradigm, we are entitled to assume that the BMV experiment will have a positive result iff  $G$ ends up in an eigenstate of $g_2$ in the course of the experiment. 

As the role of $G$ is to mediate the gravitational interaction, it must be coupled to matter in some way - i.e. $G$ must be coupled to either $Q$ or $C$. `Quantized gravity' corresponds to the case where $G$ is coupled to $Q$, whilst PIQG corresponds to the case where $G$ is coupled to $C$. We will represent this coupling by a map $M$ such that if $G$ is coupled to $X \in  \{ Q, C\}$, then $M$ takes each possible state of $X$ to a corresponding state of $G$. For simplicity we assume the map $M$ is deterministic (we don't consider that allowing a stochastic map would substantially change any of the arguments). We will assume that there is a bijection $R$ between classical mass-energy distributions and classical spacetime configurations, such that if $G$ is coupled to $X \in \{ Q, C\}$  and $X$ is in an eigenstate $d$ of an observable which commutes with $q_1$ then $G$ is in the state   $R(d)$ which is an eigenstate of an operator which commutes with $g_1$. The reason this map is a bijection is that according to the Einstein Field Equations there exists (at least locally and subject to a choice of manifold topology)  a one-to-one map between mass-energy configurations and spacetime geometries, and thus we expect this feature to be carried over to the relationship between classical variables and classical  configurations of the gravitational field. Thus the map $M$ reduces to the map $R$ for the special case where $X$ is in an eigenstate which represents  a classical mass-energy distribution.

\subsection{Inference One \label{inf1}} 

Suppose that the quantum description of the massive objects is complete, so we have no nonquantum sector.  Therefore $G$ must be coupled to $Q$. Putting a massive object in a spatial superposition corresponds to preparing $Q$ in an eigenstate of $q_2$, so we need to decide how the map $M$ acts on eigenstates of $q_2$. There are two options:   either $M$ maps  eigenstates of $q_2$ to eigenstates of $g_1$, or $M$ maps  eigenstates of $q_2$ to   to eigenstates of $g_2$, which does not commute with $g_1$. The latter option entails that there will be a superposition of gravitational fields, and thus according to our assumptions the BMV experiment will have a positive result.  The former option does not lead to any superposition of gravitational fields, but note that since each eigenstate of $g_1$ is associated by the bijection $R$ with an eigenstate of $q_1$, the map $M$ from eigenstates of $q_2$ to eigenstates of $g_1$ can be regarded as a map which takes eigenstates of $q_2$ to eigenstates of $q_1$ and then applies the map $R$ to arrive at eigenstates of $g_1$. For example, one obvious way to implement this is to take the expectation value of the quantum stress-energy tensor at each spacetime point, producing a classical stress-energy tensor defined on a single spacetime to which the Einstein tensor can be coupled; this is the approach taken by semiclassical gravity, which we discuss further in section \ref{semi}. So by going down this route we have effectively added to our quantum description an additional beable $x$ which is always in an eigenstate of $q_1$ and thus has no noncommuting variables; that is to say,  this route is tantamount to adding a nonquantum sector into our description, in contradiction with the assumption that the quantum description is complete.  

Of course, given that in this picture the classical variable $x$ is determined by the quantum state, in principle proponents of the $\psi$-complete approach would be free to continue insisting that only the quantum state is real  and  $x$ is a just a  mathematical construction which doesn't correspond to a `beable' or an element of the ontology. However,  under the circumstances this would seem to stretch credulity: after all $x$ has clear and direct physical significance since $G$ is coupled directly to it, and in that sense it has a more direct connection to macroscopic reality than even the quantum state itself. Thus a model of this kind would seem most naturally understood as a form of PIQG. Indeed if some model of this kind could be made to work, one might be tempted to conclude that only the beable $x$ is physically real, with the quantum state being regarded as modal or nomic  as in refs \cite{inbooksahc}\cite[p. 10--11]{Durr}. Moreover, the existence of this physically significant classical quantity $x$   would automatically provide us with a solution to the measurement problem, because we would be able to have classical reality supervene on the classical variable $x$ rather than directly on the quantum sector, thus ensuring that we never end up with macroscopic superpositions.  So it would seem somewhat perverse to insist that a picture such as this is genuinely $\psi$-complete. And therefore it is reasonable to conclude that if the quantum  description of the massive objects is complete in a meaningful sense, the spatial superposition of the massive objects will most likely be accompanied by a superposition of gravitational fields and hence the BMV experiment will have a positive result.

\subsection{Inference Two} 

Suppose that the BMV experiment has a positive result, and hence from our assumptions it follows that the superposition of massive objects  gives rise to a superposition of gravitational fields,  i.e. $G$ ends up in an eigenstate of a variable $g_2$ which does not commute with $g_1$. We take it that $G$ is coupled to  $X \in \{ Q, C\}$. Since $M$ reduces to $R$ in the case where $X$ is in an eigenstate of an operator which commutes with $q_1$, we know that if $X$ is in an eigenstate of an operator which commutes with $q_1$ then $G$ must be in an eigenstate of an operator which commutes with $g_1$: thus in this scenario $X$ cannot be in an eigenstate of an operator which commutes with $q_1$. But the only possible states of $C$ are eigenstates of $c_1$ which commute with $q_1$. Hence $G$ must be coupled to $Q$ rather than $C$, and therefore the quantum sector must be physically real  - where  we assume that any acceptable understanding of the term `physically real' will have the consequence that  if gravity can couple to something then it is   physically real. Therefore it is reasonable to conclude that if this experiment has a positive result, then the quantum sector is physically real.

\subsection{Inferences Three and Four} 
 Since 3) is the contrapositive of 1) and 4) is the contrapositive of 2), similar arguments can be made  in these cases. 

\subsection{Inference Five} 

For this inference we need two additional assumptions. First, we need the stronger version of the tripartite paradigm - in particular, we need to assume that the gravitational field is to be identified with the metric structure of spacetime, so superpositions of gravitational fields are also superpositions of different spacetimes. The second assumption we require is as follows:

\begin{definition} 
\textbf{Continuity:}  if  there is a nonquantum sector $C$ associated with a mass in superposition, then the nonquantum beable(s) must be in a definite state of the variable $c_1$ in the spacetime region occupied by the superposition
\end{definition} 

The continuity assumption means that we are focusing on interpretations like de Broglie-Bohm - one of the defining features of the de Broglie-Bohm particles is that they are supposed to travel on continuous paths through spacetime, so it would seem that if there exist spacetime superpositions then we must be able to say something about where the dBB particles go in the region occupied by the spacetime superposition\footnote{Tthere exists a version of Bohmian quantum gravity which potentially allows the Bohmians to get around this problem and predict a positive result to the BMV experiment; we will discuss this in more detail in section \ref{changegr}, but for now we will simply note that this approach is not a counterexample to inference five,  since this model disrupts the relationship between matter and spacetime in a way which makes it inconsistent with the strong tripartite paradigm.}. More generally,   a state of $c_1$ represents a classical mass-energy distribution, so it must be defined on some spacetime -   and which spacetime  could this distribution possibly be defined on in the region in which we have a spacetime superposition?  It can't be defined on just one of the  branches of the spacetime superposition, as then $C$ would not be in a definite state of $c_1$ - rather it would be in a superposition of a classical mass-energy configuration in one branch and no mass-energy at all in the other branches. But we also can't define the distribution on more than one of the spacetimes, as then we could in principle produce different mass-energy configurations in each branch, so $C$ would again fail to be in a definite state of $c_1$.   The problems get even worse if we argue from universal coupling that the nonquantum beables $C$ must also be coupled directly to $G$, because then we also have to decide which of the spacetimes in the superposition feels the backreaction from the nonquantum degrees of freedom. So it seems very natural to conclude that if there are spacetime superpositions then all the material content of reality (or at least all the material content of the region occupied by the spacetime superposition) must enter into superpositions, with particles, fields and so on all defined separately within each branch of the spacetime superposition, and thus all this material content must be quantum in the sense of possessing noncommuting variables. Thus since the strong tripartite paradigm tells us that a positive result to the BMV experiment entails the existence of a superposition of spacetimes, it follows that if we  assume continuity, a positive result  entails that all matter is quantum and hence the quantum description is complete. 

We reinforce that the problem of saying what happens to the nonquantum sector in the region of a spacetime superposition is a serious one for proponents of $\psi$-incomplete models, because typically the whole purpose of  introducing a nonquantum sector is to be able to say that reality, and/or our conscious experiences, supervene directly on the nonquantum sector and not on the quantum sector, which provides a straightforward explanation for the fact that we experience the macroscopic world as being in a single definite state at all times. For example, Bell writes of the de Broglie-Bohm interpretation that `\emph{in the pilot wave picture ... the wavefunction ... manifests itself to us only by its influence on the complementary variables}.'\cite{jumpsBell} But if we allow the nonquantum sector to be defined separately on two different spacetimes which are in a spacetime superposition, then we are right back where we started - we still need an explanation for why we always experience a single definite reality even though there exist spacetime regions where the nonquantum stuff on which reality and/or our experiences supervene  is doing different things in different branches of the superposition. Either we must find a way of extracting definite experiences out of the superposition, in which case it would seem that there was no need to introduce the nonquantum stuff to account for definite experiences  in the first place, or we must postulate some mechanism such as a gravitationally induced collapse which ensures that such spacetime superpositions never become large enough to have a noticeable effect on our conscious experiences, in which case it would seem that this mechanism is in itself enough to guarantee definite experiences, so again we have no need for the nonquantum sector. Therefore it is crucial to interpretations of this type that we avoid spacetime superpositions in order that the nonquantum stuff can be in a definite state at all times.

There are of course ways of out of this problem if we are prepared to deny the assumption of continuity. For example, we could  postulate nonquantum beables which don't need to have locations during the time of the spacetime superposition. One way to do this would be to invoke a picture like the Bell flash ontology, where  the nonquantum beables are `collapses' or `flashes' which mark the end of superpositions, and thus by definition are not defined within the region occupied by the superposition.  So the existence of spacetime superpositions does not wholly rule out the possbility of $\psi$-incomplete models. Nonetheless,   these considerations do point to  a notable tension between the existence of spacetime superpositions and the existence of nonquantum beables, so it does seem   that if there are physically real spacetime superpositions, postulating a nonquantum sector becomes significantly less attractive. Thus  if tabletop experiments are regarded as providing reliable evidence of the existence of spacetime superpositions, we can regard that as additional evidence for a $\psi$-complete approach.

\subsection{Inference six} 

The assumption needed for this inference can also be understood as following from the strong tripartite paradigm. In particular, since  the strong tripartite paradigm   includes commitments to all of  the  key  features of GR's account of spacetime, it should also be committed to universal coupling, i.e. the property that all matter fields couple in the same way to the gravitational field\cite{Will_2006}. Specifically, we take it that this means  that if $Q$ is physically real, it must be coupled to $G$.  
 
Moreover, as noted in section \ref{inf1}, if $Q$ is coupled to $G$ in such a way that eigenstates of $q_2$ are mapped by $M$ to eigenstates of $g_1$, then we have effectively added to our quantum description an additional beable $x$ which is always in an eigenstate of $q_1$. And as we argued above, under such circumstances we have compelling reasons to regard $x$ as an element of reality in its own right. So if $Q$ were coupled to $G$ in such a way, we would ultimately have to say that $G$ is really coupled to $x$ rather than $Q$; that is, we would be dealing with a PIQG model and $Q$ would in fact fail to be coupled to $G$ at all, violating universal coupling. So if we insist that $Q$ must be coupled directly to  $G$, it must be the case that eigenstates of $q_2$ are mapped by $M$ to eigenstates of a variable $g_2$ which does not commute with $g_1$, meaning that we will get superpositions of spacetimes and the BMV experiment will have a positive result; so it seems reasonable to think that if the quantum sector is physically real, then spacetime superpositions should be possible, and the BMV experiment will have a positive result.  Thus if we consistently fail to produce spacetime superpositions in tabletop experiments, we may wish to reconsider the notion that the quantum sector is physically real in any strong sense.

\subsection{Summary}

It follows from these inferences that if we reliably find that the BMV experiment and other similar experiments have a negative result, we will have fairly strong reasons to conclude that the quantum state description is not complete, and thus to adopt a   $\psi$-incomplete interpretation. Conversely, if these experiments  reliably have a positive result, we will have good reasons to conclude that the quantum sector is physically real, and thus to rule out $\psi$-nonphysical interpretations. Hence the results of these experiments will give us important new information about the interpretation of quantum mechanics.

Indeed, if we accept the assumptions going into inferences five and six then we can treat tabletop experiments straightforwardly as a probe of the reality of the quantum sector - if we consistently see a positive result in the BMV experiments, that should strengthen our conviction that the quantum sector, and perhaps \emph{only} the quantum sector, is  physically real (ie. this result should increase our degree of belief in $\psi$-ontic and particularly $\psi$-complete interpretations), whereas if we persistently fail to see a positive result in the BMV experiment, that should strengthen our conviction that the nonquantum sector, and perhaps \emph{only} the nonquantum sector, is physically real (i.e. this result should increase our degree of belief in $\psi$-incomplete and particularly $\psi$-nonphysical interpretations). This conclusion is of course subject to all of the caveats we have already given, and so  a positive result in these experiments would not completely rule out the possibility of $\psi$-incomplete and $\psi$-nonphysical models, but even taking these considerations into account we contend that the results of these experiments will certainly be a very important piece of new evidence for the debate over the reality of the quantum state.

A possibility we have not yet discussed is that these tabletop experiments will fail because it is simply not possible to put masses above some size in spatial superpositions. For example, this is predicted by Penrose's gravitational collapse approach\cite{Penrose1996}. This result would also have consequences for the interpretation of quantum mechanics - at first we might simply be inclined to put the failure to produce these superpositions down to engineering deficiencies, but if no progress were made and no other explanation were forthcoming then we would have good reason to take seriously either Penrose's model or some other model which predicts that superpositions which are too large will inevitably collapse. Note that Penrose seems to accept that superpositions of gravitational fields and hence spacetime structures are in  principle possible on a very small scale, though they will immediately collapse when they get big enough; this is different to PIQG, where spacetime superpositions simply can't occur on any scale because spacetime is coupled only to nonquantum degrees of freedom. In principle it should be possible to distinguish experimentally between these possibilities: we need only attempt to put a large mass in a spatial superposition, then look for signature behaviours which verify that a) we have succeeded in creating the spatial superposition (e.g. using interference as a witness) and b) we have succeeded in producing a superposition of different gravitational fields (e.g. using the BMV effect as a witness). If we obtain results indicating the mass is indeed in a superposition but we then fail to see any of the behaviours that witness the presence of a   superposition of gravitational fields, we may be inclined to conclude that   superpositions of gravitational fields are not possible,  whereas if we never manage to see the results indicating that we have successfully put the mass in a superposition, we might be more inclined to conclude that superpositions of gravitational fields may be possible in principle but some sort of collapse mechanism prevents them from becoming too large.

 \section{PIQG \label{PIQG}}
 
The literature contains a number of arguments purporting to show that gravity must be quantized, and furthermore models where gravity is quantized more or less universally predict the existence of  superpositions of gravitational fields and/or spacetimes. Thus one might be tempted to argue that we already know  that superpositions of gravitational fields and/or spacetimes must exist, and so if we fail to detect them in tabletop experiments, the only possible conclusion is that our experimental techniques are not currently good enough. This line of argument would suggest that PIQG models have already been ruled out and therefore no interesting possibilities remain to be ruled out by the tabletop experiments. 

However, all of the arguments for the quantization of gravity require certain assumptions: for example Tilloy points out that many of them rest `\emph{on a dangerous straw man: taking flawed mean-field semiclassical approaches as representatives of all hybrid quantum-classical theories.}'\cite{2018bqmast} So while these arguments do help to narrow down what a theory of quantum gravity without quantized gravity must look like, they don't actually rule such a thing out. The flaws of these arguments have been examined in some detail in refs  \cite{2018bqmast,Craig2001-CRAPMP,inbookmatt}, so we will not dwell on them here - in appendix   \ref{mustbe} we examine these arguments in the context of PIQG and demonstrate that none of them actually rules out PIQG models, but for now let us pass to an examination of specific PIQG proposals.  

In this section we first summarise some relevant proposals; we then discuss what these approaches have in common and what would need  to be done to make the PIQG approach more broadly appealing to physicists. It's not our intention to argue that any existing PIQG model is at present a serious contender for the correct theory of quantum gravity,  but we hope to show that this general space of models remains viable given current empirical evidence and hence  if tabletop gravity experiments do in fact rule out PIQG models, that will  be an important and meaningful result.

\subsection{Semiclassical gravity \label{semi}} 

The most well-known approach to quantum gravity without spacetime superpositions is semi-classical gravity, which is based on   Rosenfeld’s 1963 suggestion\cite{1963NucPh..40..353R} that the classical Einstein tensor could be set proportional to the expectation value of the quantum stress-energy operator. To do this, we construct a quantum field theory on a curved spacetime and then allow the stress-energy tensor to couple to the Einstein tensor via what is known as the semiclassical Einstein equation:

\[ G_{uv} = \frac{8 \pi G} {c^4} \langle T_{uv} \rangle_{\psi} \] 

Now one may well feel it is somehow unsatisfying to have a fundamental equation in which an expectation value appears, and thus the semiclassical equation is usually regarded as being only an approximation which is to be understood as arising from some underlying theory, such as low energy quantum gravity, and moreover it can be extended to a better approximation known as stochastic semiclassical gravity which aims to account for fluctuations around the expectation value of the stress-energy tensor in a self-consistent way\cite{Hu_2008}. However, one could in principle imagine regarding  semiclassical gravity as a fundamental theory in and of itself. Superficially this doesn't look like a  PIQG model since it does  not   explicitly employ a nonquantum sector; but  as  argued in section \ref{inf1} this approach is tantamount to introducing a nonquantum sector, since the expectation value $\langle T_{uv} \rangle_{\psi}$, which has no noncommuting variables, plays a central role in the theory and therefore has a reasonable claim to be regarded as physically real in its own right. So semiclassical gravity may be regarded as a simple example of  a PIQG model. Moreover, in the case where  semi-classical gravity is treated as an approximation, it is possible to imagine that the semiclassical Einstein equation might arise from  an underlying theory in which gravity is not quantized, such as a PIQG model. After all, in any model employing a stress-energy tensor which supervenes entirely on a nonquantum sector, it seems likely that the value of that  stress-energy tensor would at least in some cases be  fairly close to $\langle T_{uv} \rangle_{\psi}$, since the distribution of the nonquantum sector in $\psi$-incomplete models is usually closely related to the distribution defined by the mod-squared amplitude of the wavefunction. Indeed, an empirically successful PIQG model would certainly have to approximate semiclassical gravity in certain cases, for semiclassical gravity makes accurate predictions in several easily observable regimes\cite{wallace2021quantum}. Therefore semiclassical gravity may have important lessons for PIQG models.

That said, ideally   PIQG models should be formulated in ways that avoid some of the obvious obstacles encountered by the semi-classical gravity programme. First, one serious problem for the semiclassical approach is that  it predicts `gravitational cat states' for macroscopic matter distributions. That is, in a version of the semiclassical theory which does not employ a collapse mechanism, if we manage to create a macroscopic superposition of a particle in two locations, then the gravitational field will show peaks in both of the possible location. But applying the standard Born rule interpretation would lead us to expect that we will always find the gravitational field to be peaked in exactly one location, and  indeed an experiment due to Page and Geilker\cite{PhysRevLett.47.979} is usually regarded as having verified that prediction, so the semiclassical theory seems to be empirically inadequate.  That said,   the Page and Geilker experiment verifies this prediction only if we assume that the quantum state always evolves unitarily\cite{wallace2021quantum}, so arguably semi-classical gravity in combination with a dynamical collapse model may still be viable. But either way, these kinds of experiments make it clear that   a successful PIQG model must have the property that even if the wavefunction shows  a superposition of a particle in two locations, for sufficiently large particles the matter making up the nonquantum sector will be concentrated  in only one of those locations, so the gravitational field predicted by the PIQG model will show a peak in only one location and will therefore be consistent with the empirical data. Refs \cite{2020sca, 2014maan} exhibit PIQG models (discussed in  sections \ref{dBB} and \ref{GRW} below) which implement this property and thus succeed in avoiding gravitational cat states. 
 
Second,  it has been noted that  $\langle T_{uv} \rangle_{\psi}$   has ultra-violet divergences arising from the mathematically ill-defined short-distance behaviour of quantum fields, and regularization methods only yield an unambiguous expression when the spacetime metric is time-independent, which will not usually be the case\cite{butterfield_isham_2001}; so semi-classical gravity can't currently be rigorously defined. There are also problems arising from the fact that some quantum states allow   $\langle T_{uv} \rangle_{\psi}$  to be negative, which doesn't make much sense in the equation above as $G_{uv}$ can't be negative. And ref  \cite{butterfield_isham_2001}  also notes that it often unclear how to choose the state $| \psi \rangle$ to calculate $\langle T_{uv} \rangle_{\psi}$  in the first place. These considerations suggest that a successful PIQG model will most likely not use the quantum stress-energy tensor directly - instead it will employ a stress-energy tensor defined directly on the distribution of nonquantum matter. For example, if we take it that the stress-energy  tensor supervenes on something like the density of flashes in the Bell flash ontology, then since the flashes are discrete and there is nothing in between them, we will never be forced to deal with short-distance behaviour. Similarly, it is clear that a tensor which is defined in terms of the density of flashes will never become negative.

Third, in the non-relativistic limit the semiclassical gravity equation can be used to derive a non-linear evolution equation known as the Schr\"{o}dinger-Newton equation, and the non-linearity of this equation is known to give rise to further problems\cite{Tilloydoweneed}. In particular,  the equation couples  orthogonal branches of the wavefunction, meaning that decoherence can no longer be relied upon to prevent interactions between macroscopically distinct branches of the wavefunction. This is a serious problem for any interpretation of quantum mechanics which does not postulate wavefunction collapses, including most $\psi$-incomplete models, because it will lead to predictions which are clearly wrong - for example it would lead us to predict an attraction between the dead and alive branches of Schrodinger's cat\cite{Tilloydoweneed}. This suggests that a successful PIQG model should probably yield a linear evolution equation in the non-relativistic limit. As demonstrated by Tilloy's GRW flash model\cite{2018grw} (discussed in section \ref{GRW}) such a thing is certainly possible.

\subsection{Bohmian approaches \label{dBB}} 

Bohmian mechanics\cite{Passon2006-PASWYA,holland1995quantum,SEPBohm} is a $\psi$-incomplete interpretation of quantum mecahnics which solves the measurement problem by introducing an actual configuration of $N$ particles, with the configuration  $(Q_1, Q_2 ... Q_N)$ guided through spacetime by the wavefunction according to the guiding equation $\frac{d Q_k} {dt} = \frac{\hbar}{m_k }  Im (\frac{\psi^* \partial_k \psi }{ \psi^* \psi }  ( Q_1, Q_2 ... Q_N))$. Reality, and/or conscious experience, is understood to supervene on the actual configuration and not the wavefunction, so all measurements have unique outcomes and there is no possibility of encountering macroscopic superpositions. Some versions of Bohmian mechanics regard the wavefunction as a real physical field, in which case the interpretation is $\psi$-supplemented, whereas some versions of Bohmian mechanics read the wavefunction as something like a law of nature, in which case the interpretation is $\psi$-nonphysical. 

One obvious possibility for a Bohmian PIQG model would be to let the stress-energy tensor supervene entirely on the configuration of the particles,  then since the particles always have a single actual configuration, no superpositions of spacetimes need arise. This approach has  been proposed  as a method of arriving at better semiclassical approximations\cite{2020sca} -  in particular, it avoids gravitational cat states, since the Bohmian theory guarantees that in superpositions of macroscopically different possibilities, the Bohmian particles will  necessarily be associated with only one branch. However, in principle one could imagine regarding this as a candidate fundamental theory in which we always have a single classical spacetime sourced by a the Bohmian particles. One difficulty for such an approach  is that it would fail to be relativistically covariant, since the original Bohmian model without gravity does not satisfy Lorentz covariance due to the fact that the `configuration' of particles featuring in the Bohmian guidance equation given above is defined on a single time-slice, and thus  we have to choose a preferred foliation on which the evolution takes place. In the non-gravitational version the preferred foliation turns out to be undetectable because there are limitations on the degree to which we can get to know the positions of the Bohmian particles, but if the particles were to source a gravitational field this would offer new opportunities for gaining information about their positions and thus there would be a danger that the preferred reference frame might become observable: for example, as noted by Callender and Huggett, `\emph{Signalling could therefore occur if scattering at the gravitational field depended on the particle configuration and not only the wave function}.'\cite{Craig2001-CRAPMP} 

This indicates that even if we're willing to acccept the failure of Lorentz covariance at the level of the undetectable parts of the  model, in order to be empirically adequate a successful Bohmian model of this kind would have to be written in such a way that  we can't  use the coupling between the Bohmian particles and the structure of spacetime to get to know the positions of the  particles with precision greater than the probabilistic description given by the wavefunction. One way to achieve this would be to insist that the relationship between the value of the stress-energy tensor and the Bohmian variables should be a little `fuzzy'; another possibility would be to insist that there are limits on the precision with which we can measure the curvature of spacetime. Note that we seem to require that the amount of information we can gain about the Bohmian variables corresponds \emph{exactly} to the wavefunction probability distribution: too much information and we will violate no-signalling, too little and the model fails to be empirically adequate, since in real life we can at least sometimes figure out the wavefunction. So one might worry that it seems suspiciously coincidental that the amount we can learn about the Bohmian variables from their coupling to spacetime corresponds so precisely to the wavefunction probability distribution, even though the wavefunction is supposedly not involved in this coupling.

As we will discuss in section \ref{changegr}, there exists a different approach to Bohmian quantum gravity in which the 3-metric itself is a Bohmian variable, but this is not a PIQG model since it tells us that spacetime is not directly to coupled to matter.

\subsection{Collapse-based approaches \label{GRW}}

The GRW collapse model\cite{GRW,Frigg2009} is a $\psi$-incomplete interpretation of quantum mechanics which solves the measurement problem by postulating that quantum systems undergo spontaneous collapse events which cause their wavefunction to become strongly peaked at a single spacetime location. When one system undergoes a spontaneous collapse all systems entangled with that system will also undergo a collapse. Thus we can choose the probability distribution   for the collapses  such that it is relatively unlikely for an individual quantum state to collapse, but overwhelmingly likely for a macroscopic object composed of many entangled particles to undergo a collapse in any given interval;  so macroscopic objects will always be found in definite states, and since measurements outcomes are recorded on macroscopic devices, measurements will always have unique outcomes. There are several different ontologies which can be associated with the GRW collapse model\cite{2020po}. In the mass-density approach\cite{Allori_2013}, we postulate that the wavefunction describes a distribution of mass across spacetime, with the mass at a given point $x$ being proportional to the value of $|\psi(x)|^2$, so there are instantaneous changes in the mass distribution every time a collapse occurs. Alternatively in the Bell flash approach\cite{Bell1985,Gisin2013}, the quantum state is regarded as purely nomic,  leaving us with an ontology composed only of the collapse events, or `flashes.'

A non-relativistic version of semiclassical gravity based on the mass density version of the GRW collapse approach is given in ref \cite{2014maan}.  Here we simply calculate the gravitational potential felt by a particle at a given point by integrating over the mass density proportional to $|\psi(x)|^2$, so the gravitational potential is always single-valued at any spacetime point. However, this approach is not Lorentz covariant, for  if the collapse instantaneously localises all the mass at the collapse centre, then  the mass must instantaneously disappear from everywhere else, and that process must occur on some preferred reference frame. It follows that an extension of this approach to the context of general relativity would most likely fail to be generally covariant. Moreover, if we used the mass density to define an energy-momentum tensor, the covariant divergence of this energy-momentum tensor would necessarily be nonzero during a wavefunction collapse, whereas the covariant divergence of the Einstein tensor is always zero, so it would not be consistent to set this energy-momentum tensor equal to the Einstein tensor.  

Alternatively, we can employ the flash ontology. As noted earlier,  the flash ontology tells us that there is actually no matter present in regions which we describe in terms of superpositions, and thus for the proponent of the flash ontology it actually seems most natural to predict that there are no spacetime superpositions.   A model of this kind has been postulated by Tilloy\cite{2018grw}, with each flash being `smeared' across spacetime according to some smearing function, creating a mass distribution which sources the gravitational field. Formally speaking the flashes are measurement outcomes, and therefore we can model the  effect of the gravitational field sourced by a flash on the ongoing quantum state just as we would in a feedback scheme where an observer performs a measurement and then applies an operation conditioned on the measurement result:  `\emph{immediately after a flash, an instantaneous unitary transformation is applied to the wave-function corresponding to the singular gravitational pull of the flash.}' As   noted by Tilloy, this has the notable advantage that it can all be achieved within the standard quantum formalism and thus everything remains linear.  

 Note that in general the smearing  will single out a preferred reference frame and hence for most smearing functions this model will not exhibit Lorentz covariance; however, if we take the limit as the smearing goes to zero, the potential created by the flash becomes divergent but the unitary operation modelling the effect of the flash on the wavefunction at the time of the flash will remain finite, so this may offer a route to Lorentz covariance. That said,  because Tilloy's approach involves a unitary operation applied immediately after a flash, and these operations will not in general commute, it follows that we  have to think of the flashes as being generated in some temporal order, and that order will single out a preferred reference frame: as observed by Esfeld and Gisin, the Bell flash ontology is relativistically covariant only if we `\emph{renounce any account of the temporal coming-into-being of the flashes}'\cite{Gisin2013}. So it would seem that Tilloy's model in its current form will not exhibit Lorentz covariance.

\subsection{Looking Forward}

The obvious flaw shared by these accounts is that they don't have the right covariance properties - most of them fail to satisfy Lorentz covariance, and thus when extended to a general relativistic setting they would most likely fail to be generally covariant\footnote{As noted by Kretschmann, nearly any theory can be put in a `generally covariant' form such that the theory's equations do not change when the spacetime coordinates are transformed. However, there are stronger formulations of general covariance, such as the requirement  subsequently suggested by Einstein that a theory's equations should take their simplest form when they are written in a generally covariant language\cite{https://doi.org/10.1002/andp.19183600402}, or the requirement that the generally covariant formulation of the theory does not introduce new auxiliary quantities which fail to be physically motivated\cite{fock2015theory}. The general relativistic version of a theory which fails to satisfy Lorentz covariance will in general fail to satisfy these stronger general covariance principles.}  While it remains technically possible that the actual universe fails to have these properties provided that the failures are  not observable in the kinds of experiments we have so far performed, few physicists find that possibility appealing. So if PIQG is to have wider appeal, it is important to come up with a PIQG approach which exhibits Lorentz covariance and thus has an appealing formulation in a generally covariant language. 

It's not particularly surprising that the models above fail to exhibit the desired covariance properties, because they are based on interpretations of quantum mechanics which have problems with Lorentz covariance even in the absence of gravity. For example, it is well-known that the de Broglie-Bohm approach in its standard formulation postulates an instantaneous state update everywhere which must take place on a preferred reference frame\cite{1997ogal}; likewise, a collapse model with a mass-density ontology must   postulate a preferred reference frame on which the mass-density distribution undergoes a discontinuous change as the wavefunction collapses\cite{2020po}. The Bell flash ontology is supposed to be Lorentz-covariant, but this holds only if we refrain from assigning  an overall temporal order to the flashes\cite{Gisin2013}, so inevitably an approach to PIQG based on a flash model with some temporal ordering will not be Lorentz invariant. 

So an obvious way to make progress would be attempt to formulate PIQG using some interpretation of quantum mechanics which at least satisfies Lorentz invariance in the absence of gravity.  One option would be to adopt an approach similar to Tilloy's based on the Bell flash ontology, but avoid constructing the model in a way which depends on the temporal order of the flashes.  In Tumulka's presentation of the relativistic flash approach\cite{Tumulka2006} we are provided with  with three different ways to calculate the distributions of the particles: using a foliation of spacetime into spacelike surfaces, using an iterative construction of the flashes, and using a formula for the joint distribution of the flashes on spacetime in terms of the initial wavefunction. The latter seems the most likely to work in a PIQG model - presumably the flashes and the gravitational field would be determined all-at-once in a mutually consistent way.  Another option would be to construct a PIQG model based on Kent's solution to the Lorentzian classical reality problem\cite{Kent2013,2015KentL}, an interpretation of quantum mechanics which has been explicitly designed to exhibit manifest Lorentz-invariance. In Kent's approach, the wavefunction evolves unitarily for the whole of history, and then a measurement is performed at on the resulting state at the end of time; the value of the actual stress-energy tensor at any spacetime point is then given by its expectation value conditioned on the outcome of this final measurement. Coupling the Einstein field equation to this stress-energy tensor would presumably produce something which looks a little like semiclassical quantum gravity, but some of the problems with semiclassical gravity would be solved - in particular, we would not get gravitational cat states, since the final measurement will have selected one out of the two possible spatial positions for the mass.

Another potential difficulty is that as in the Bohmian case, if the energy-momentum tensor is to supervene on `hidden variables' then we will have to be careful that the interaction between gravity and the hidden variables doesn't end up giving us too much information about the hidden variables. For example, we know from the  Colbeck-Renner\cite{RennerColbeck} theorem that if we can obtain information about these variables in such a way that we are able to use that information to make predictions more accurate than quantum mechanics, then there will be violations of no-signalling. So assuming that we don't want to violate no-signalling, PIQG models will have to   ensure that the coupling between the nonquantum sector and spacetime is sufficiently blurry to prevent us from obtaining too much information. 

However, it's important to reinforce that there are two quite different approaches to a $\psi$-incomplete model. The nonquantum sector can be made up of `hidden variables,' i.e. beables which are \emph{more} predictive than the quantum state - as for example in the de Broglie-Bohm interpretation, where perfect knowledge of particle positions would allow us to perfectly predict all measurement outcomes. But it is also possible for the nonquantum sector to be made up of beables which contain  only the kind of information which we can typically extract from the quantum state by measurement - as for example in the spontaneous collapse interpretation, where the nonquantum sector is essentially just composed of measurement outcomes (albeit  measurements which are not performed by any agent), and in Kent's Lorentzian proposal, where the beables are simply the result of a measurement performed at the end of time. In these sorts of models the information to be gained from the nonquantum beables can never be greater than the information that orthodox quantum mechanics allows us to gain by performing measurements, so  knowledge of the nonquantum sector doesn't give us better information than knowledge of the quantum state. This means that in models of this kind we don't have the added complication of needing to make some of the information in the nonquantum sector inaccessible; as such, it's likely that   PIQG approaches based on this kind of $\psi$-incomplete model are likely to be simpler. 

Finally, as noted in the discussion of section \ref{semi}, PIQG models which lead to non-linear evolution in the non-relativistic limit are likely to be inconsistent with existing empirical evidence. Fortunately there are possible ways to avoid this problem. As noted by Tilloy\cite{2018bqmast}, if the effect of gravity is modelled by a measurement followed by an operation conditioned on the measurement outcome, we stay within the quantum formalism and thus achieve linearity (the measurement itself is non-linear, but linearity is re-introduced when we average over all the measurement outcomes together with their corresponding operation). Kent's model\cite{Kent2013,2015KentL}, also offers a route to linearity, since in this picture the quantum state undergoes its usual linear evolution, only collapsing at the end of time - so provided that this linearity can be retained when a coupling to gravity is added,  one would not have to deal with the problems associated with the Schr\"{o}dinger-Newton equation in a PIQG version of this model. 

  \section{Approaches Modifying GR \label{changegr}}

The assumption made in the tripartite paradigm that  GR's account of spacetime structure is correct at the BMV scales is powerful because it allows us to make an \emph{existence} claim: a successful result  to the BMV experiments demonstrates the \emph{existence} of superpositions of gravitational fields (or superpositions of spacetimes, if we choose the strong version of the tripartite paradigm). Thus this approach gives us a direct entry into questions of ontology and allows us to draw some  strong ontological conclusions - not least because spacetime is such an central feature of reality which of course must enter in some way into every interpretation of quantum mechanics.  Conversely, if we move outside the   tripartite paradigm  where we no longer assume that the BMV experiments prove something about what kinds of superpositions can exist, it is significantly more difficult to use them as a window into quantum ontology.

That said, even if we do admit the possibility of divergences from the general relativistic picture, we can still draw certain conclusions from the results of tabletop experiments. First  off, as already noted, a positive result to the BMV experiment rules out PIQG models even if we don't presuppose that GR is correct at these scales, since the PIQG models themselves assume that GR is correct at these scales. Second, even if we allow the possibility of divergences from GR it remains the case that most $\psi$-complete models will predict a positive result to the BMV experiments, because if all of reality is quantum then gravity must be quantum in some sense and hence we would naturally expect that it would be able to exhibit quantum phenomena like interference under appropriate conditions. Conversely, we know that at least some $\psi$-incomplete models (e.g. PIQG ones)  predict that we should obtain negative results in such experiments. Hence even if we don't assume the correctness of GR at the BMV scales,  it seems reasonable to say that the \emph{failure} of the BMV experiments would be a highly significant result, causing us to have an decreased degree of belief in $\psi$-complete interpretations and an increased degree of belief in $\psi$-incomplete interpretations.  That is to say, in this setting inferences one and three remain reasonable, although we might perhaps have a little less confidence in them if we do not take the tripartite paradigm for granted. 

To illustrate some of these points, we now consider some specific examples of quantum gravity models which offer an account of spacetime and gravity significantly different from that suggested by GR.

\subsection{Bohmian Quantum Gravity \label{DBB2}} 

Although one can envision a simple Bohmian PIQG model as described in section \ref{dBB}, there exists an alternative approach to   Bohmian  quantum gravity   where we begin by choosing a foliation of spacetime and then we take the spatial 3-metric $h_{ab}$ on the slices to be a Bohmian variable   which evolves in accordance with a Bohmian guiding equation\cite{goldstein_teufel_2001,pintoneto2019bohmian,CWxx}. The Bohmian guiding equation is defined using a  wavefunction  $\Psi(h_{ab})$ over the space of possible spatial 3-metrics. This approach to Bohmian quantum gravity is of course $\psi$-incomplete, and may or may not also be $\psi$-nonphysical depending on whether    the field  $\Psi(h_{ab})$ is regarded as physical or nonphysical (e.g. it is sometimes regarded as being `nomic' rather than physical\cite{10.2307/24562842,Durr}). 

The original version of this model\cite{goldstein_teufel_2001} did not include any matter, but evidently in order to predict an outcome to the BMV experiment we must include some matter. There are two different ways we might imagine doing this. If we stick to the strong tripartite paradigm  where we take it that matter should be coupled directly to spacetime, it follows that the matter should have its own Bohmian evolution equation which is separate from the 3-metric equation, such that at all times the motion of the matter depends directly on the 3-metric $h_{ab}$ and not on any function of  $\Psi(h_{ab})$. This version of the Bohmian theory will predict a negative result to the BMV experiment, because $h_{ab}$ cannot enter into superpositions and thus regardless of the state of the matter there will only ever be one value of the gravitational force experienced on each side of the experiment, so no entanglement can possibly form. In this case, our proposed definition of `physical' would  allow us to say that the quantum sector is not physically real, since it does not directly source any gravitational fields which act on matter. 

The alternative is to define a single wavefunction $\Psi(h_{ab}, \phi)$ over the space of  joint states of the 3-metric $h_{ab}$ and the matter field $\phi$, such that the Bohmian guiding equation defined using this state guides the evolution of the real 3-metric  and the matter jointly. This version of the Bohmian theory, which seems to be the one actually advocated by proponents of Bohmian quantum gravity\cite{pintoneto2019bohmian}, predicts  a positive result to the BMV experiment  even though  $h_{ab}$ is not in a superposition, since the behaviour of the matter depends directly on the superposition of spacetimes represented in $\Psi(h_{ab}, \phi)$. However, note that the behaviour of both the matter and the 3-metric $h_{ab}$ is now determined directly by the quantum state  $\Psi(h_{ab}, \phi)$ as dictated by the Bohmian guiding equation, and therefore  the relationship between the matter and  $h_{ab}$ will always be mediated by $\Psi(h_{ab}, \phi)$. For example, the matter  responds directly to the superposition of 3-metrics which is  present in  $\Psi(h_{ab}, \phi)$ and only indirectly to the geometry of  $h_{ab}$, so the geodesics that it follows will not always be the geodesics of the spacetime  arising from the temporal development of  $h_{ab}$.  Similarly, the matter $\phi$ will not source a gravitational field: rather the part of the quantum state   $\Psi(h_{ab}, \phi)$ associated with the matter $\phi$ will be appropriately coupled to the part of the state   $\Psi(h_{ab}, \phi)$ associated with $h_{ab}$ and that in turn will determine the behaviour of the `real' 3-metric $h_{ab}$ according to the guiding equation, so $h_{ab}$ will not always be related to $\phi$ in the way that GR might lead us to expect. Indeed, $h_{ab}$ is almost entirely inert in this picture: it does not couple to matter or tell matter where to go, and we can't probe it directly using matter, as we will always find ourselves probing the  quantum state instead (this is the same mechanism as in the original Bohmian theory, where  we can't find out the exact positions of the Bohmian particles, only the quantum state to which their distribution corresponds). Thus in this model the one and only function of $h_{ab}$ is to provide a background on which the single-valued matter field $\phi$ can live.  Perhaps the correct thing to say in this situation is that `spacetime' no longer refers unambiguously to a single object, since the roles typically attributed to spacetime are now played in part by the temporal development of  $h_{ab}$ (which provides the background on which the matter lives) and the temporal development of  $\Psi(h_{ab}, \phi)$ (which takes on the dynamical properties of spacetime) so we have two different objects each taking on some of the roles of spacetime. 
 
 The fact that this Bohmian approach disagrees with GR's account of the relationship between spacetime and matter at small scales is not necessarily a problem, because the Bohmian approach will make the same predictions as classical GR in the limit where the systems involved become large, and since GR is only confirmed at relatively large scales it could certainly be the case that it is only correct in the emergent sense predicted by the Bohmian approach. That said, one might worry that there is something suspiciously coincidental about the fact that descriptions in which spacetime is coupled to matter have been so successful if we grant that in fact spacetime is not actually coupled directly to matter at all. There are also some technical problems with this Bohmian model. In particular, in canonical quantum gravity the foliation approach ultimately produces a theory which is  foliation-invariant, but this is not the case in the Bohmian version: in general the evolution of the Bohmian 3-metric over some interval will depend on the choice of intermediate hypersurfaces\cite[p. 13]{goldstein_teufel_2001}, and thus this approach to Bohmian quantum gravity  fails to be generally covariant. To solve this problem, we must either choose a preferred foliation, or place a further constraint on the universal wavefunction insisting that it must generate a foliation-independent vector field\cite[p. 13]{goldstein_teufel_2001}, but neither of these possibilities has so far been fully developed.

\subsection{EFT models}

It is common to analyse the weak gravity regime using models based on an effective field theory picture in which we have a classical  background spacetime associated with the ambient gravitational field sourced by surrounding large masses, and  then on that background we define fluctuating gravitational fields, including those sourced by small masses like the masses in the BMV experiment. As discussed in ref\cite{wallace2021quantum} this approach can be derived formally within a path integral approach, where we write the field as the sum of its expectation value plus small fluctuations; an expansion of this kind is also used in stochastic gravity, which is an extension of semiclassical gravity aiming to characterise fluctuations around the expectation value of the stress-energy tensor\cite{Hu_2008}, and in linearized gravity, which is the approximation used in refs \cite{anastopoulos2018comment,Anastopoulos_2021} to arrive at the Newtonian analysis of the BMV experiment. Thus it is clear that  an EFT approach allows us provide a description of the BMV experiment which appears to avoid postulating  superpositions of spacetimes, although it may still have superpositions of gravitational fields. 

Now, if we accept the assumptions of the strong tripartite paradigm we are compelled to regard the EFT approach as simply a convenient calculational tool, and thus we must agree that the true explanation for these phenomena will involve superpositions of spacetimes even if that is not evident in the EFT description. However, we might perhaps hope to use the EFT approach to predict a positive result to the BMV experiment in a way which accords with the (minimal) tripartite paradigm but which does not require us to postulate any spacetime superpositions.  This would involve taking the EFT description to be more than just an approximation - for example, we could take it that we really do have a classical background spacetime which does not enter into superpositions, plus a fluctuating gravitational field defined on that background which does enter into superpositions. Thus the gravitational field is still mediated, as the minimal tripartite paradigm demands, but spacetime itself remains classical.  In this sense, analysing the BMV experiment in the minimal tripartite paradigm may succeed in showing that if the BMV experiment has a positive result then some aspects of gravity must be quantum, but it does not prove that if the BMV experiment has a positive result then \emph{spacetime itself} must be quantum - in order to draw that conclusion we must advert to the strong tripartite paradigm, since then a superposition of gravitational fields implies a superposition of spacetimes. The EFT approach would also open up a route to explaining a positive result to the BMV experiments in a way that is compatible with a $\psi$-incomplete model satisfying continuity - the continuous non-quantum matter could live on the classical background spacetime whilst the superpositions of gravitational fields needed to produce the BMV effect could still be present in the same region. 

As we have noted, there are certainly legitimate reasons to think that   GR's account of spacetime might break down at small scales, so it's not impossible that spacetime and the gravitational field could come apart at the scales of the BMV experiment. However, one might worry that models in which the structure of spacetime and the gravitational field emerge separately and subsequently become unified at an even higher scale would lose something in simplicity compared to   models in which they are one and the same thing and thus emerge together. Indeed, allowing gravity and spacetime to diverge at small scales would seem to undermine the conceptual innovation of Einstein's original insight that they are one and the same thing, so those who find Einstein's unification appealing have good reason to resist this kind of strategy.  

At any rate, as things stand the approach which treats the gravitational field and the background spacetime as separate entities is not fully satisfactory as an account of the nature of the gravitational interaction. For a start, it would need  to be supplemented with some model explaining what happens in the limit as the superposed masses become large. Moreover, given that the   choice of decomposition into `background field' and `fluctuating field' is arbitrary, some justification is certainly required before we can make a deep ontological distinction  such that the background field is identified with spacetime and the fluctuating field is not identified with spacetime: writing the field as the sum of its expectation value plus small corrections is mathematically convenient, but this does not in and of itself provide any reason to think the expectation value part is qualitatively different from the small corrections. 

In light of this one might perhaps  appeal to an approach like that of Knox\cite{KNOX2013346}  in which spacetime geometry is regarded as  a  high-level feature which need not be precisely defined at small scales,  so we are free to say that  there is no need to choose any particular decomposition into background field and fluctuating field, since what counts as `spacetime' is defined only approximately. However, even if we find this account satisfactory in general,    if we want to use the effective field theory approach to advocate for a $\psi$-incomplete picture something more will be needed, as we will have to explain why and how the non-quantum matter responds differently to the ambient gravitational field and the superposed quantum fluctuations, and this would presumably require us to specify some preferred EFT decomposition. One way this might be achieved would be to derive the EFT description from some underlying theory (e.g. an emergent gravity model) which has the consequence that `\emph{gravitational low energies degrees of freedom are qualitatively different from the high-energy degrees of freedoms (for example if the spin-2 particle is a composite particle)}'\cite{LinnemannVisser}. In such a model, the background field and the quantum fluctuations would genuinely be different types of objects and thus the EFT decomposition would no longer be ad hoc; however, it remains to be seen if this can be done in a compelling way.

\subsection{Emergent Gravity} 

The Bohmian  and  EFT approaches depart from GR in certain ways but they do  maintain something which looks like the standard GR coupling between spacetime and matter, even if the coupling is indirect or only holds in certain regimes. More radically,    there exist alternative approaches known as `emergent gravity' which tell us that GR's account of the nature of gravity must ultimately be replaced with something very different: as defined by Linnemann and Visser\cite[p. 8]{LinnemannVisser}, these are models in which GR is derived from an underlying microtheory which is not directly inspired or motivated from GR, so by construction  these theories tell an ontological story which is very different from that of GR. If we allow for possibilities which depart from our established understanding of spacetime in this radical way, it's hard to say anything concrete  about what we can conclude from the tabletop experiments: this is essentially an instance of the problem of unconceived alternatives\cite{stanford2006exceeding}. Thus, although   these approaches are very interesting, it may not be very informative to try to analyse the tabletop experiments in the context of this very general space of possibilities - the only feasible way to proceed is to consider what specific emergent gravity approaches have to say about the tabletop experiments (see ref \cite[p. 13]{LinnemannVisser}  for a comprehensive taxonomy of existing possibilities).
 
For example, in Sakharov's induced gravity model, gravity is not fundamental but is instead induced by quantum field theory\cite{Sakharov:1967pk}. In this approach, one does quantum field theory on a Lorentzian manifold, and then it turns out that `\emph{the effective action  at the one-loop level automatically contains terms proportional to the cosmological constant and to the Einstein-Hilbert action of general relativity}'\cite{doi:10.1086/508946}. It is not clear whether this kind of approach is adequate to fully recover gravity, and the use of a fixed background may perturb proponents of general covariance, but nonetheless Sakharov's approach does demonstrate that there exist viable alternatives to the story that GR has to tell about the relationship between matter and spacetime. In particular,  Sakharov's model demonstrates another way in which the link between a positive outcome to the BMV experiment and the existence of spacetime superpositions could fail if we are outside the (strong) tripartite paradigm, because this model does not involve superpositions of spacetimes but it seems very possible that a fully developed version of this model would, if interpreted as $\psi$-complete,  still predict a positive result for the BMV experiment. That said, the model could presumably also be combined with some other interpretation of quantum mechanics, such as a gravitational collapse model or similar,  and in that case it would not necessarily predict a positive result for the BMV experiment. In this sense, even though the induced gravity approach is very far removed from the tripartite paradigm, nonetheless its predictions for the BMV experiment are still closely related to a choice of interpretation of quantum mechanics.

\section{The Cosmological Constant \label{cosmology}} 

We have focused in this article on tabletop experiments, but it's also worth noting that there are other domains where different approaches to interpreting quantum mechanics would be likely to make broadly different predictions. In particular, it has long been noted that cosmological phenomena are likely to be good places to look for evidence of quantum gravity phenomena, and since we have seen in this article that different interpretations tend to favour different approaches to quantum gravity, such cosmological effects may also be regarded as indirect tests of the interpretation of quantum mechanics. 

Indeed, there have already been proposals for specific cosmological predictions which would follow from some of the alternative quantum gravity approaches we have discussed in this article - for example, ref \cite{pintoneto2019bohmian} discusses the consequences of Bohmian quantum gravity (in the non-PIQG version of section \ref{DBB2}) for a variety of cosmological phenomena.  However, these predictions largely depend on quite specific features of individual models, none of which we can presently have a high degree of confidence about; so it would be interesting to see if there are any cosmological phenomena which are similar to spacetime superpositions in the sense that  $\psi$-complete models naturally lead to one kind of prediction and $\psi$-incomplete models naturally lead to some other kind of prediction. This would enable us to draw  interpretational conclusions from cosmological phenomena which don't depend on the details of any specific model. 

For example,  many physicists are particularly concerned about what is known as the `cosmological constant problem' - i.e. the fact that the observed value of the cosmological constant is much lower than predicted by low-energy quantum gravity based on the assumption that vacuum fluctuations gravitate\cite{doi:10.1119/1.17850, wallace2021quantum}. As noted in ref \cite{RUGH2002663} one obvious way to account for this observation would be to  say that vacuum fluctuations are not physically real; or of course, alternatively one might say that vacuum fluctuations are physically real but they do not gravitate. Now  in a $\psi$-complete context it would seem difficult to come up with a good reason why vacuum fluctuations should not gravitate\footnote{Unless we are willing to get rid of the standard formalism of QFT and replace it with something like Schwinger's source theory\cite{Schwinger:1970xc}}: we don't really have the option of saying something like `one should not associate energy to fields in empty space,' as suggested in ref \cite{RUGH2002663}, because if matter consists entirely of fields then \emph{all} fields are fields in empty space, so this stipulation would prevent us from associating energy to anything at all. 
 
But in a $\psi$-incomplete context we have the option of supposing that the energy-momentum tensor supervenes entirely on the nonquantum sector and that furthermore the beables of the nonquantum sector occur only where there is matter, not in the vacuum, so vacuum fluctuations don't  gravitate. The dialectic here is  similar to the spacetime superposition case: in a $\psi$-complete model it seems natural to suppose that vacuum fluctuations gravitate,  whereas in a $\psi$-incomplete model there's a feasible route which allows us to say that vacuum fluctuations don't gravitate, although we are not necessarily compelled to take that route. Insofar as this argument is accepted it would seem that the low value of the cosmological constant might be regarded as evidence for $\psi$-incomplete models, though of course in order to make this argument really compelling it would be necessary to formulate some specific $\psi$-incomplete model which predicts a value for the cosmological constant consistent with current observations but also manages to reproduce the successes of existing cosmological models. 

Naturally, there are a number of obstacles in the way of such a project. In particular, current cosmological models tell us that in the very early universe quantum fluctuations magnified by gravitational effects during inflation  were responsible for the formation of structures like galaxies and galaxy clusters\cite{1967ApJ...147...73S, wallace2021quantum}, so one would either have to undertake the daunting task of finding another way of accounting for structure formation which reproduces the empirical success  of the existing model, or  find a way of arguing that those early fluctuations contained gravitating nonquantum beables even though most vacuum fluctuations currently do not. One way to make the latter argument would be by appeal to Kent's final-measurement interpretation. For Kent's interpretation has the consequence that the distribution of nonquantum beables is determined by the result of the final measurement, and thus beables can be associated with some event only if there remains a record of that event in the quantum state at the end of time. Clearly   it is the case that early-universe vacuum fluctuations have left stable ongoing records, since the `structures' they form act as records of them which are still around today, and thus according to Kent's approach they will have beables associated with them;  whereas most current vacuum fluctuations do nothing interesting and leave no ongoing records, and thus according to Kent's approach they will typically fail to have any beables associated with them. Therefore if we take it that a PIQG generalisation of Kent's model would have the consequence that only the nonquantum beables defined by the final measurement gravitate, it would follow that early-universe quantum fluctuations should gravitate but most  current vacuum fluctuations would not gravitate, thus accounting for the low value of the cosmological constant. 

Another difficulty is that there are a variety of  well-confirmed physical phenomena which are typically attributed to vacuum fluctuations: for example, the experimentally demonstrated Casimir effect is typically regarded as a consequence of zero-point energies\cite{2004casimir}. So if we  want to insist that vacuum fluctuations don't gravitate because they don't contain any physical beables, we would   have to  find a way of reproducing the  behaviours typically attributed to the vacuum fluctuations in the absence of any physically real vacuum fluctuations. But in fact this is just a rehashing of the same objections that have often been made against $\psi$-nonphysical (or $\psi$-epistemic) interpretations in ordinary non-relativistic quantum mechanics: if the wavefunction is not real, how do we get interference, entanglement, and various other behaviours that are typically explained in terms of the  wavefunction? Thus a PIQG approach could draw on the same kinds of solutions as ordinary $\psi$-nonphysical models do: quantum fields, and vacuum fluctuations in particular, may be regarded as a form of modal structure or a manifestation of the laws of nature, so they are capable of having physical effects (just as other sorts of modal relations and laws of nature have physical effects) but they do not gravitate (because modal structures and laws of nature don't have spacetime locations and don't carry mass or energy). We hope to examine these possibilities in greater detail in future work.

\section{Conclusion}

In this paper we have argued that, in the context of the tripartite analysis of the BMV experiments, $\psi$-complete models generally predict that the BMV experiment will have a positive result;   $\psi$-nonphysical models often predict that the BMV experiment will not have a positive result; and for $\psi$-supplemented models there may be arguments for either possibility, although it does seem hard to combine the requirement that there is a  nonquantum sector made up of entities travelling along continuous paths in spacetime   with the existence of spacetime superpositions. Thus insofar as the results of tabletop experiments for quantum gravity give us information about whether or not gravity is quantum, these experiments will also give us information about the correct interpretation of quantum mechanics. In our opinion, this is an exciting moment for the foundations community - while the results of the tabletop experiments won't conclusively disprove any interpretations, they will certainly be a very interesting piece of new evidence and it is our hope that this may breathe new life into the longstanding impasse over the interpretation of quantum mechanics. 

Moreover, these considerations aptly demonstrate the deficiencies of the common claim that research on the interpretation of quantum mechanics is pointless because  all interpretations of quantum mechanics make the same predictions. For the real test of these interpretations will come precisely when we extend them into other domains, most obviously quantum gravity: given any interpretation there  are more and less natural ways of extending it into the gravitational regime, and if experiments agree with the predictions of the most natural extension of some particular interpretation, that will  certainly be a strong point of evidence in the interpretation's favour. This underlines the fact that  theory selection in science is seldom based on absolutely conclusive verification or falsification: in most cases theories become accepted not via a single conclusive test but  because of the way they fit into a bigger picture, including considerations about the way in which they interact with theories of other regimes and the extent to which they inspire new developments. Thus new experimental data such as the results of the tabletop experiments is certainly capable providing empirical evidence that favours certain interpretations of quantum mechanics over others. Of course this may not be the kind of irrefutable proof  that people presumably have in mind when they complain that interpretations of quantum mechanics are untestable, but to ask for a conclusive test is to misunderstand the nature of scientific confirmation: in all likelihood the measurement problem will ultimately be resolved not via a single critical experiment but by means of the accumulation of theoretical insights and  new experimental data which over time point more and more strongly toward some particular interpretative approach. 

At this juncture one might object that it is fruitless to make interpretational speculations  about quantum mechanics at all until we have a final theory of quantum gravity, since we already know that quantum mechanics can't be the most fundamental theory. And we do agree that the sorts of arguments we have made in this article will most likely look rather naive in hindsight, as it is very likely that the final theory of quantum gravity will look nothing at all  like the simple heuristic picture we have considered here with a gravitational field coupling either to a quantum or nonquantum sector. However,   we have seen in this article that answers to interpretational questions bear on  the approach one takes to quantum gravity in the first place, so if we simply ignore these questions until we have a satisfactory theory of quantum gravity, there's a danger of missing the correct route to quantum gravity altogether. Thus in reality the dialogue needs to go both ways:  attempts to interpret quantum mechanics should take heed of ongoing research on quantum gravity, but also quantum gravity researchers should be conscious of their implicit interpretational commitments and  possible alternatives. Thus the results of upcoming tabletop experiments will offer a fruitful opportunity for the two fields to come together and hopefully inspire new progress on both sides of this question.

\section{Acknowledgements} 

 Thanks to Niels Linnemann, Nick Huggett, Mike Schneider and the UWO philosophy of physics reading group for very helpful discussions. This publication was made possible through the support of the ID 61466 grant from the John Templeton Foundation, as part of the “The Quantum Information Structure of Spacetime (QISS)” Project (qiss.fr). The opinions expressed in this publication are those of the author  and do not necessarily reflect the views of the John Templeton Foundation.

\appendix

\section{Arguments that gravity must be quantum \label{mustbe}}

\subsection{Eppley and Hannah} 

An argument due to Eppley and Hannah dating to 1977 \cite{Eppley1977-EPPTNO} purports to show that the gravitational field must be quantized. Eppley and Hannah consider the interaction of a classical gravitational wave of small momentum with a quantum particle described by a wave function. If we suppose that this interaction does not collapse the wavefunction whereas ordinary measurements \emph{do} collapse the wavefunction, it follows that this setup can be used to perform superluminal signalling, as  the wavefunction of both particles will collapse when Bob performs a measurement, and Alice can then use the gravitational wave to probe the state of her particle to see whether Bob has performed his measurement or not. But if we suppose this interaction does collapse the wavefunction, then Eppley and Hannah argue that there will be problems with momentum conservation or violations of the Heisenberg uncertainty relation. So assuming that we are not willing to accept violations of no-signalling, momentum conservation of the Heisenberg uncertainty relation, it follows that the gravitational field must be quantized. 

As noted by Huggett and Callender\cite{Craig2001-CRAPMP},  the first horn of the dilemma makes interpretational assumptions by  assuming that nongravitational  measurements lead to a wavefunction collapse; evidently interpretations which don't tell us that measurements cause wavefunction collapses won't have this problem. In particular, $\psi$-incomplete models don't typically postulate a collapse of the wavefunction upon measurement, because definite classical results for measurements are already assured by the fact that measurement results supervene directly on the nonquantum sector, so there's no need for a wavefunction collapse. For example, the Bell flash ontology does postulate `flashes' which are roughly similar to collapses, but they are not \emph{caused} by measurement: they're spontaneous and uncontrollable and thus can't be used for signalling. Moreover,  in  a PIQG model  spacetime is  coupled  to the nonquantum sector rather than the quantum sector, so gravitational waves can't directly probe the quantum sector. Thus as long as the nonquantum sector is defined in such a way that the knowledge Alice can gain about the nonquantum sector via a gravitational probe doesn't give her too much information about the quantum sector, no violation of signalling  will ever occur. 

The second horn of the dilemma is also suspect -  as Mattingly points out `\emph{it may be that the uncertainty relations \emph{can} be violated. They haven’t really been tested in this way}.'\cite{inbookmatt} In any case this horn is not really relevant to PIQG models, as if gravity is coupled directly to the nonquantum sector there is never any direct interaction between gravity and the quantum sector, so presumably the wavefunction will not be collapsed by gravity.

Therefore the argument of Eppley and Hannah does give us valuable information - it tells us that in formulating a theory of quantum gravity in which gravity is not quantized, we should probably avoid approaches where measurements and/or gravitational interactions cause  a collapse of the wavefunction - but it certainly doesn't rule out PIQG models.

\subsection{DeWitt}

An argument due to DeWitt dating to 1962\cite{PhysRev.125.2189} argues that the quantization of any system, as expressed by the uncertainty principle, implies the quantization of all other systems to which it can be coupled. This is a generalization of an argument used by Bohr and Rosenfeld to show that the electromagnetic field must be quantized\cite{BRaf}. However, the argument relies on understanding the uncertainty principle in terms of `disturbance,' and it has since been recognised that this approach is not correct\cite{Brown1981-BROACO-7}.  And in any case, in a PIQG model gravity is not coupled to the quantum sector, so DeWitt's argument does not apply. Of course, in a PIQG model typically we will have the quantum sector coupled to the nonquantum sector, so one might try to invoke DeWitt's argument twice over to show first that the nonquantum sector must be quantized and then that gravity must be quantized. However, it is clearly not the case that a nonquantum sector coupled to a quantum sector must be quantized - numerous counterexamples exist, including the de Broglie Bohm interpretation\cite{Passon2006-PASWYA,holland1995quantum,SEPBohm}. DeWitt's arguments do not apply to this kind of case, because he proceeds by supposing that the measurement of a system $A$ by another quantized system $B$ will create a disturbance which leads to $A$ becoming quantized, whereas the coupling between the de Broglie-Bohm particles and the quantum state does not proceed by measurement and hence is not subject to uncertainty relations.

\subsection{Marletto and Vedral}

Recently Marletto and Vedral presented a novel information-theoretic argument for the quantization of gravity\cite{marletto2017need}. The aim here is to show that if a  system $C$ is coupled to a quantum system $Q$ with two non-commuting variables, then $C$ must also have at least two non-commuting variables. Rather than making assumptions about the dynamics of the coupling as in previous proofs, Marletto and Vedral  work entirely with information-theoretic concepts within the constructor theory approach. In particular, they assume `interoperability' (i.e. that classical information on $Q$ can be copied to $C$) and they  make an assumption which we will refer to as $A*$, which requires that  an operation which measures the variable $x_1$ on $Q$ and then copies the result to $C$ will also be a `distinguisher' for the non-commuting variable $x_2$ on $Q$, i.e. it will map the possible values of $Q$ to states of $C, Q$ which are perfectly distinguishable. 

The assumption $A*$ may seem quite specific, but Marletto and Vedral justify it by the claim that it is true in the case of quantum mechanics. And indeed, this is correct if   $C$ and $Q$ are both quantum systems. For example, suppose $Q$ is a qubit and $C$ is a second qubit which we denote by $Q'$. Now if $Q$ is prepared in the computational basis (variable $x_1$) and $Q'$ is prepared in the state $|0 \rangle_{Q'}$, we can use a CNOT with $Q$ as the control to `measure' $Q$ and copy the result to $Q'$. Meanwhile if $Q$ is prepared in the Hadamard basis (variable $x_2$) and $Q'$ is again prepared in the state $|0 \rangle_{Q'}$, and we apply a CNOT with $Q$ as the control, if $Q$ was in state $| + \rangle_{Q}$ the result will be the Bell state $\frac{1}{\sqrt{2}} ( | 00 \rangle_{QQ'} + |11 \rangle_{QQ'})$, whereas if $Q$ was in the state $| - \rangle_{Q}$ the result will be  the Bell state $\frac{1}{\sqrt{2}} ( | 00 \rangle_{QQ'} - |11 \rangle_{QQ'})$, so the two possible values of $x_2$ are indeed perfectly distinguished. 

However, what we actually want is for $C$ to be the gravitational field, so in this example it is necessary that the state of the second qubit $Q'$ should be mapped to a state of the gravitational field. One option is to simply measure $Q'$ in the computational basis after applying the CNOT gate; then if the result is $0$ we move a large mass to the left (producing state $|L \rangle_M$) and if the result is $1$ we move a large mass to the right (producing state $| R \rangle_M$).  Thus in the case where we are measuring the variable $x_1$, the final state of the gravitational field can be regarded as a record of the result of the measurement of $x_1$. Now let us consider what happens if instead we attempt to perform the $x_2$ distinguishing operation. The argument of Marletto and Vedral requires that we perform exactly the same operations as in the case where we are measuring $x_1$, so we must apply the CNOT gate and then measure $Q'$ in the computational basis and then move a large mass around conditioned on the result of that measurement. However, according to the usual interpretation of measurement, measuring the second qubit destroys the coherence between the qubits; therefore we will just end up with either state the state $|0 \rangle_{Q'} | L \rangle_{M}$ or the state $|1 \rangle_{Q'} | R \rangle_{M}$ with fifty percent probability for each regardless of whether $Q$ was originally in state $|+ \rangle_{Q}$ or $| - \rangle_{Q}$, meaning that the distinguishing property is lost.  Of course, we could retain the distinguishing property if we instead measured the pair of qubits $Q, Q'$ in the Bell basis and then moved around the mass conditional on the results of that measurement; but the argument of Marletto and Vedral does not allow that, as we are required to use the same operation in both cases. Thus the assumption $A*$ does not hold if the intention is for us to map the qubit state to the gravitational field by means of measurement. 

The other option is to make both of the qubits $Q,Q'$ massive objects, with state $|0\rangle$ corresponding to one spatial position and the state $| 1 \rangle$ corresponding to a different spatial position, such that $C$ can be dentified with the gravitational field sourced by these massive objects. Then, if it is the case that spatial superpositions of massive objects give rise to superpositions of spacetimes, it follows that the  two Bell states $\frac{1}{\sqrt{2}} ( | 00 \rangle_{QQ'} + |11 \rangle_{QQ'})$,   $\frac{1}{\sqrt{2}} ( | 00 \rangle_{QQ'} - |11 \rangle_{QQ'})$ will give rise to different spacetime superpositions, thus retaining the coherence and preserving the distinguishing property on which the argument relies. But this begs the question: whether or not massive objects in superpositions produce spacetime superpositions is precisely the point at issue, so we can't simply assume that the distinguishing property is retained when information is mapped to a state of the gravitational field. 

Here we have considered just one possible implementation of the scenario envisioned by Marletto and Vedral, but it seems likely that   the same kinds of problems will hold for all possible implementation of the scenario. For according to the textbook account of quantum measurement, a coupling  implemented by measuring qubits is capable only of copying information in a single basis, so we cannot have a coupling  of this kind which preserves both the computational basis states for the $x_1$ measurement and also the Bell states for the $x_2$ distinguishing operation. So assumption $A*$ will hold only if the coupling is achieved by some kind of quantum operation which transfers the whole quantum state of the qubits into a state of the gravitational field, which is possible only if the gravitational field is itself quantum and spacetime superpositions are possible. Thus $A*$ does not hold in any PIQG model,  since such models do not allow the existence of spacetime superpositions. 
 
 Note that Marletto and Vedral might respond to this objection by rejecting the idea that measurement destroys coherence and instead adopting an Everettian picture. In that case, the result of measuring the qubit and transferring the result to the gravitational field would not in principle destroy the distinguishability of the two Bell states, although they would now only be distinguishable in an abstract external sense, since no person within any individual branch of the wavefunction would be able to distinguish them. So the argument of Marletto and Vedral arguably does succeed in an Everettian context - but then this is not very surprising, as we have already noted that the existence of spacetime superpositions seems more or less inevitable in an Everettian picture.

\subsection{Belenchia et al} 

Ref \cite{2018qsom} provides an interesting new argument for the quantisation of gravity. They consider a case where one particle $A$ is able to obtain information about the spatial position of another particle $B$ by probing $B$'s gravitational field; then if $B$ is in a superposition of spacetime positions, if follows that if $A$ is present the state of $B$ must become less pure as $A$ becomes entangled with it, whereas if $A$ is not present the state of $B$ remains pure, and thus we seem to have a case of superluminal signalling. The resolution to this apparent paradox is to perform a more careful analysis which takes account of vacuum fluctuations and the quantum properties of radiation. Thus it is  argued that gravity must at the last  possess the properties of a quantum field with regard to vacuum fluctuations and the quantum properties of radiation, so it must be quantum. 

Now, this is not an argument for quantisation per se - rather it is an argument to the effect that if gravity can be used to produce entanglement as in the BMV experiment, then it must indeed have quantum features.  In the scenario described in the article, a PIQG model would simply deny that $A$ could become entangled at all: for example, a PIQG model might hold that there are actually no beables  present in the region where $B$ is in a spacetime superposition, so  $B$ does not source any gravitational fields during this time, so $A$ can't obtain any information about its position from the gravitational field and no paradox can arise.  So our the approach of ref  \cite{2018qsom} is not at all inconsistent with our approach in this article - indeed, this argument may be understood as supporting our choice to assume the correctness of the tripartite paradigm in our analyses in section \ref{inference}.

That said,   the argument   isn't completely conclusive. For what has been shown is simply that in the case where gravity \emph{is} a quantum field, the resolution to the paradox is to consider vacuum fluctuations and the quantum properties of radiation. It doesn't necessarily follow that there is no possible resolution to the paradox if gravity is not quantum; it would simply have to be a \emph{different} resolution. However, insofar as no classical resolution is currently available, and a well-motivated quantum resolution is available, this argument does seem to offer solid reasons to adopt the quantum approach. Moreover, ref \cite{2018qsom} notes that a very similar paradox can be set up using electromagnetic rather than gravitational interactions, and it would seem somewhat unsatisfying  to have one resolution of the paradox in the case where the interaction is electromagnetic and a completely different resolution in the case where the interaction is gravitational: if the electromagnetic and gravitational interactions lead to exactly the same kinds of behaviour, the natural conclusion is that the two interactions are also of the same kind, i.e. they are both quantum. PIQG models would of course deny that electromagnetic and gravitational interactions do lead to exactly the same kinds of behaviour, since PIQG models do not allow superposed particles to source gravitational fields, but this argument is primarily directed at the case where superposed particles \emph{can} gravitate, and from that standpoint it does indeed paint a convincing picture to the effect that that if superposed particles gravitate, gravity must possess the properties of a quantum field.

\subsection{Unification}

As  noted in section \ref{intro}, arguments for the quantisation of gravity frequently appeal to the need for unification. Of course one possible rebuttal would be to  argue for a pluralist or instrumentalist approach which denies that unification is always desirable or necessary, but there is no need for us to make that argument here, because we consider that PIQG models  do in fact count as a form of unification. 

In particular, we observe that `unification' can mean several different things. Unifying theories from two different regimes may sometimes take the form of reducing one theory to another, or reducing them both to a single deeper underlying theory. These sorts of unifications typically lead us to conclude that the objects postulated by the theories are in fact one and the same kind of object - for example, showing that the gravitational field is in fact a quantum field. But sometimes unification may just involve giving a consistent account of the relationship between two theories which explains how the different types of objects that they postulate interact with one another and how transitions between the two regimes work. If we are committed to the former kind of unification then it may seem reasonable to conclude that the gravitational field must be quantum (although of course we could also choose to insist that all of physical reality is \emph{non-quantum}), but if we are committed only to the latter kind of unification then there is nothing compelling us to say that the gravitational field must be quantum. 

Moreover, although there are certain sorts of metaphysical commitments that might lead someone to prefer the former sort of unification,   the latter is perfectly adequate if our interest in unification simply stems from the simple realist position that ultimately there exists a single universe and so  all the parts of that universe must fit together into one consistent schema. Nothing about this entails that there can only be one kind of of `physical stuff' - of course it may be easier to figure out how to fit everything together into a consistent schema if we have only one type of stuff, but the universe is not obliged to make things easy for us, and therefore the conjecture that there is only one type of stuff is subject to confirmation or disconfirmation by empirical data just like any other hypothesis. The preference for only one type of stuff seems in some cases to be almost an aesthetic preference, but of course if the empirical data strongly indicated that there is more than one kind of physical stuff then it would be bad practice to disregard that evidence on purely aesthetic grounds.  At present the empirical data does not seem wholly conclusive, but convincingly demonstrating either the existence or the nonexistence of spacetime superpositions would be a major step towards reaching a conclusion on this point. 

Tilloy also mentions  a related kind of argument\cite{2018bqmast}: the enthusiasm for quantizing gravity  stems in part from the hope that successfully quantizing gravity will also solve some of our other theoretical problems - e.g. it may regularize the UV divergences of Quantum Field Theory and tame the singularities of General Relativity. But as Tilloy points out, these hopes have not yet been realised, so at present these arguments look a lot like wishful thinking. Moreover, one might well feel that there is just as good reason to think that finding the correct solution to the measurement problem will solve these problems, particularly if it turns out that solution is not $\psi$-complete; so it does not seem methodologically sound to insist that quantizing gravity is necessary in order to solve these problems.

\subsection{Success of QG so far} 

An argument that does not appear much in the literature but which is often brought up informally  begins from the observation that all currently existing  well-developed approaches to quantum gravity postulate spacetime superpositions, and proceeds to the conclusion that we have good reason to believe in spacetime superpositions. Indeed, given the vast size of the literature on string theory\cite{Blau2009} and the rapidly increasing literature on Loop Quantum Gravity\cite{cc}, it may seem tempting to conclude that we could not have made so much theoretical progress on quantizing gravity if quantizing gravity were not in some sense the right thing to do. 

We do agree that the progress that has been made on these approaches into account is significant and should be assigned due weight when making assessments about the quantization of gravity.  On the other hand, theoretical progress is not the same as empirical confirmation: after all mathematicians routinely discover all sorts of interesting structure associated with  abstract mathematical objects which as far as we know don't represent anything in reality, so the mere fact that we have discovered interesting structure within theories of quantized gravity doesn't in and of itself entail that they represent anything in reality. Thus until empirical confirmation becomes available it would be unwise to shut off all other possible routes. 

Moreover, most well-developed approaches to QG - including canonical quantum gravity, LQC and string theory - took as their starting point the need to quantize the gravitational field, and therefore it is no coincidence that they predict spacetime superpositions. So the theoretical success of these research programmes does not necessarily give us reason to believe in spacetime superpositions unless we are convinced that similarly serious attempts have been made to formulate a theory of interactions between quantum systems and gravity \emph{without} quantizing the gravitational field, and that those attempts have failed. It doesn't seem that this is the case: throughout the history of the field, `quantizing gravity' has largely been the favoured approach, whereas semiclassical gravity and PIQG have received comparatively little attention. Of course, as we discuss in section \ref{semi}, there are a number of very good reasons to be sceptical about semiclassical gravity, but nonetheless the fact that non-quantized gravity approaches have received significantly less energy and attention makes it hard to draw any strong conclusions from the fact that quantized gravity has made more theoretical progress than non-quantized gravity.

A  different version of this argument would appeal specifically to the success of low energy quantum gravity. For unlike string theory, LQC and so on, low energy quantum gravity is quite well confirmed\cite{wallace2021quantum}, and thus it might seem reasonable to extrapolate from the success of low energy quantum gravity in certain regimes to the overall correctness of the theory. Since   low energy quantum gravity assumes that gravity is quantized and that spacetime superpositions can exist, this would entail that gravity must indeed be quantized. That said, the theory of `low energy quantum gravity plus dynamical or gravitational collapse' makes roughly the same predictions as standard low energy quantum gravity in the regimes that we have tested but diverges in the so-far untested regime where superpositions of spacetimes become insignificant, so the empirical evidence does not obviously distinguish between these two alternatives. Similarly, the Bohmian PIQG approach discussed in section \ref{dBB} can reproduce much of the success of low energy quantum gravity, and indeed it seems to give rise to better semiclassical approximations in certain regimes\cite{2020sca}, so the evidence doesn't seem to rule Bohmian views out either. Thus the success of low energy quantum gravity doesn't rule out PIQG approaches, because most PIQG approaches would be expected to coincide with low energy quantum gravity in the regimes in which it has been tested.

 \bibliographystyle{unsrt}
 \bibliography{newlibrary12}{}

\end{document}